\documentclass[acmsmall]{acmart}

\usepackage{graphics} 
\usepackage{epsfig} 
\usepackage{gensymb}
\usepackage{subfig}
\usepackage[export]{adjustbox}
\usepackage{paralist}
\usepackage{commath}

\usepackage{color}
\usepackage{booktabs}
\usepackage{flushend}
\usepackage{graphicx}
\usepackage{textcomp}
\usepackage{xcolor}
\usepackage{multirow}
\usepackage{longtable}

\graphicspath{{Figures/}}
\DeclareGraphicsExtensions{.pdf,.png,.jpg,.eps} 

\begin{document}

\def\acmBooktitle#1{\gdef\@acmBooktitle{#1}}
\acmBooktitle{ACM Journal \acmConference@name
       \ifx\acmConference@name\acmConference@shortname\else
         \ (\acmConference@shortname)\fi}

\title{Circuit Masking: From Theory to Standardization} 
\subtitle{A Comprehensive Survey for Hardware Security Researchers and Practitioners}


\author{Ana Covic}
\email{anaswim@ufl.edu}
\affiliation{$^1$}
\author{Fatemeh Ganji}
\email{fganji@wpi.edu}
\affiliation{$^2$}
\author{Domenic Forte}
\renewcommand{\shortauthors}{Covic, et al.}
\email{dforte@ece.ufl.edu}
\affiliation{$^1$}
\affiliation{%
  \institution{\\
  1 Florida Institute for Cybersecurity Research, University of Florida}
  \streetaddress{601 Gale Lemerand Dr}
 \city{Gainesville}
  \state{FL}
 \postcode{32611}}
 \affiliation{
  \institution{2 Worcester Polytechnic Institute}
  \streetaddress{100 Institute Road}
 \city{Worcester}
  \state{MA}
 \postcode{01609-2280}
}



\begin{abstract}
Side-channel attacks extracting sensitive data from implementations have been considered a major threat to the security of cryptographic schemes. 
This has elevated the need for improved designs by embodying countermeasures, with masking being the most prominent example. 
To formally verify the security of a masking scheme, numerous attack models have been developed to capture the physical properties of the information leakage as well as the capabilities of the adversary. 
With regard to these models, extensive research has been performed to realize masking schemes. 
These research efforts have led to significant progress in the development of security assessment methodologies and further initiated standardization activities.
However, since the majority of this work is theoretical, it is challenging for the more practice-oriented hardware security community to fully grasp and contribute to. 
To bridge the gap, these advancements are reviewed and discussed in this survey, mainly from the perspective of hardware security. 
In doing so, a clear taxonomy is provided that is helpful for a systematic treatment of the masking-related topics. 
By giving an extensive overview of the existing methods, this survey (1) provides a research landscape of circuit masking for newcomers to the field, (2) offers guidelines on which attack model and verification tool to choose when designing masking schemes, and (3) identifies interesting new research directions where masking models and assessment tools can be applied. 
Thus, this survey serves as an essential reference for hardware security practitioners interested in the theory behind masking techniques, the tools useful to verify the security of masked circuits, and their potential applications.

\end{abstract}



\maketitle

\section{Introduction}
\label{sec:intro}
The theory of cryptography usually considers the interaction between the adversary and a cryptosystem in an abstract setting, sometimes called a ``black-box.'' 
In this regard, it is assumed that the only information correlated with the secret key available to the adversary is limited to pre-defined interfaces, i.e., inputs and outputs of the system. 
In practice, however, the adversaries can obtain information about the secret key and the internal state of computation as the cryptographic algorithm is ultimately implemented on a physical platform~\cite{kalai2019survey}. 
This indeed has a measurable impact on the environment around the platform; for instance, the execution time, power consumption, and electromagnetic emanation (so-called side channels) can leak key-dependent information. 
This type of leakage, referred to as computation leakage, considers the scenario, where the side-channel information comes from the intermediate values during the computation, rather than only from the secret itself~\cite{kalai2019survey}
A large variety of successful side-channel attacks leveraging this information have been reported in the literature. 
Designing countermeasures against such attacks has motivated a line of research focusing mainly on ``masking.''

Circuit masking is a provably secure technique developed for protection against probing and side-channel attacks on integrated circuit (IC) chips. 
As an effective and theoretically sound countermeasure, masking is built on the pioneering work of Chari et al.~\cite{chari1999towards}, and Goubin et al.~\cite{goubin1999and}. 
These studies have independently suggested to apply the so-called XOR-secret sharing or Boolean masking countermeasure: each bit $b$ is represented by $k$ random bits, whose exclusive-or is equal to $b$.  
To implement this in practice, true-random number (TRN) generators (TRNGs) are used to, roughly speaking, introduce confusion to the original circuit's functionality. The inputs of the circuit are \textit{masked} with the TRN, while each circuit's gate is replaced by a cluster of gates called \textit{gadget}, which has the same functionality as the original gate, but it is augmented by TRNs. 

Similar to other cryptographic primitives, the first step towards designing a masked circuit is the definition of an attack model that can capture the characteristics of the leakage and the attacker's capabilities. 
The next step includes designing building blocks known as ``gadgets'' that fulfill the necessary requirements of the attack model, which need to further be compatible with a composable security model. 
This is due to the fact that even though the gadgets are secure, their composition may be insecure. 
Although effective in the pre-silicon stage, this methodology should be evaluated when the physical implementation of the scheme is available. 
This is necessary even when the attack models take into consideration physical effects, namely glitches and coupling~\cite{buhan2021sok}. 
In fact, the development of practical circuit masking schemes poses increasingly difficult challenges for designers as the above-mentioned steps must be followed by security verification in the post-silicon phase, which may indicate that the pre-silicon steps should be repeated to achieve the desired level of security. 
This is despite the efforts made to propose adequate security verification methods and tools that even have pushed the development of attack models and gadgets. 

The ongoing research on the pre-silicon and post-silicon steps involved in the design cycle of masking schemes has attracted the attention of researchers from academia, industry, and standardization bodies. 
For instance, the National Institute of Standards and Technology
(NIST) has initiated the Threshold Cryptography project investigating the side-channel resistance~\cite{NIST2020}.
Similarly, the European Commission is coordinating a framework~\cite{EU2020}, including projects dealing with the
implementation of leakage detection tools, see, e.g.,~\cite{reassure2020}. In line with this, protection against side-channel
attacks is among the criteria defined by certification bodies in several countries.


Nevertheless, we believe that there is a lot of room for improvement because there is a lack of understanding of the outstanding work done by the cryptographic community from the hardware security community. The majority of the cryptography community's work has been theoretical, even when considering physical implementations. Consequently, the side of the hardware security community that is more practice-oriented is not aware of the theoretical work already done. 
This survey aims to bring this elegant work performed by the cryptography community closer to the hardware security community by describing the notions from a more practical perspective. Our main contributions are as follows: 
\begin{itemize}
    \item Succinct explanations, descriptions, and definitions of probing models and corresponding gadgets, various types of theoretical masking types, and masking implementations
    \item Summary of security verification frameworks and tools
    \item Review of standardization efforts
    \item Future directions for development of masking schemes and their use in other problems relevant to the hardware security community,
\end{itemize}

\begin{figure}
  \includegraphics[width=\linewidth]{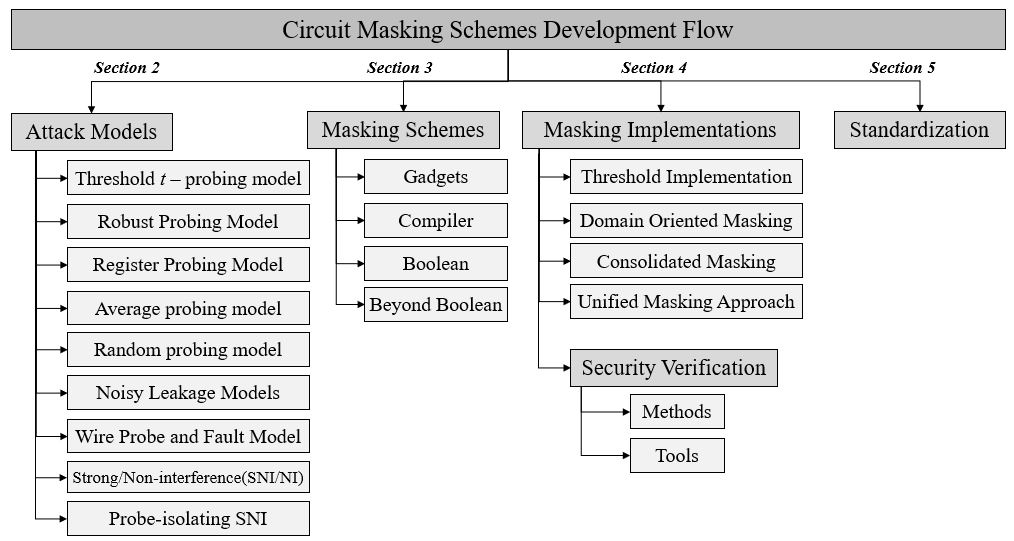}
  \caption{A taxonomy for circuit masking schemes research and development}
  \label{fig:tax}
\end{figure}

It is worth noting the extensive surveys carried out by Kalai et al.~\cite{kalai2019survey} and Buhan et al.~\cite{buhan2021sok} with the aim of increasing the knowledge within and outside the hardware security community. 
The former has focused on more theoretical findings within the scope of ``leakage-resilient cryptography,'' whereas Buhan et al. have shifted their focus to a practical aspect: available tools for leakage detection and security verification. 
Compared to these studies, our survey can be seen as an introduction of both theoretical and practical aspects to hardware security practitioners. 
This survey starts with a description of \textit{Attack Models} in Section~\ref{sec:attack_models} where we introduce probing models, noisy leakage models, security notions, and fault-injection attack models. In Section~\ref{sec:masking_schemes}, we introduce various gadgets, as well as the importance and use of randomness. In that section, there is also a discussion and summary of various masking techniques. In Section~\ref{sec:masking_implementations}, we introduce various implementation schemes and corresponding gadgets, as well as the security verification techniques and tools. In Section~\ref{sec:standardization}, we bring standardization efforts, as shown in Figure~\ref{fig:tax}. 
In Section~\ref{sec:lessons} we discuss the future directions of hardware security community using the knowledge learned from this survey, as shown in Figure~\ref{fig:tax}. 
Finally, we conclude in Section~\ref{sec:conclusion}.

\section{Attack Models}
\label{sec:attack_models}


To develop countermeasures that protect circuits against various probing and side-channel attacks, the security must be verified against those attacks~\cite{danger2018physical}. Consequently, theoretical attack models have been established which describe the abilities of an attacker. Designing countermeasures based on circuit masking consists of several steps. As seen in Figure~\ref{fig:tax}, the first step is to identify the theoretical attack model. Next, for the masking scheme, building blocks referred to as \textit{gadgets} must be developed. 

\begin{figure}
  \includegraphics[width=0.5\linewidth]{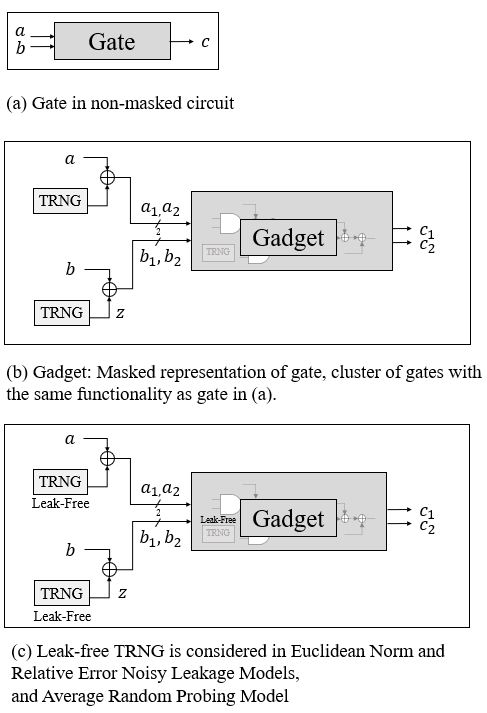}
  \caption{Input(share)-output dependency in gadget with $t=2$ shares. A gadget is a masked representation of the simple gate. Each input and output needs to be split into shares. For the case of 2 shares, an attacker needs to have $t+1$ (3) probes to extract any meaningful information.}
  \label{fig:gadget1probe}
\end{figure}

A gadget is a cluster of gates arranged in a specific way to have the same functionality as its corresponding original gate, but further requires that each bit of original gate inputs be split into \textit{shares}. This is done by introducing random numbers from a random number generator, where the randomness of the gadget input shares, as well as internal randomness, can be described as the ``freshness'' of the gadget inputs or signal. Correspondingly, when all input shares are fresh (random), and the number of the shares is $t+1$, the attacker cannot extract information from the output of the gadget with $t$ probes. For example, for the case of two shares per input, as in Figure~\ref{fig:gadget1probe}, an attacker needs to have three probes to extract information about shares. Three probes are needed to extract information because of the two shares per input. To extract any meaningful information, the attacker needs to have one probe more than there are the shares per input. This means that there are more probes than random input values, creating the possibility to extract information from its correlation. 

Once the gadgets are determined to be secure under the theoretical attack model, an entire original circuit (consisting of standard gates) can be transformed into a masked circuit (consisting of gadgets). Note that many early-developed gadgets do not provide sufficient security. It has been shown that once they are composed together on the circuit level, they are susceptible to physical hardware defaults (e.g., glitches). Thus, it has been a goal of the community to improve composition and masking schemes. 

\begin{figure}
  \includegraphics[width=\linewidth]{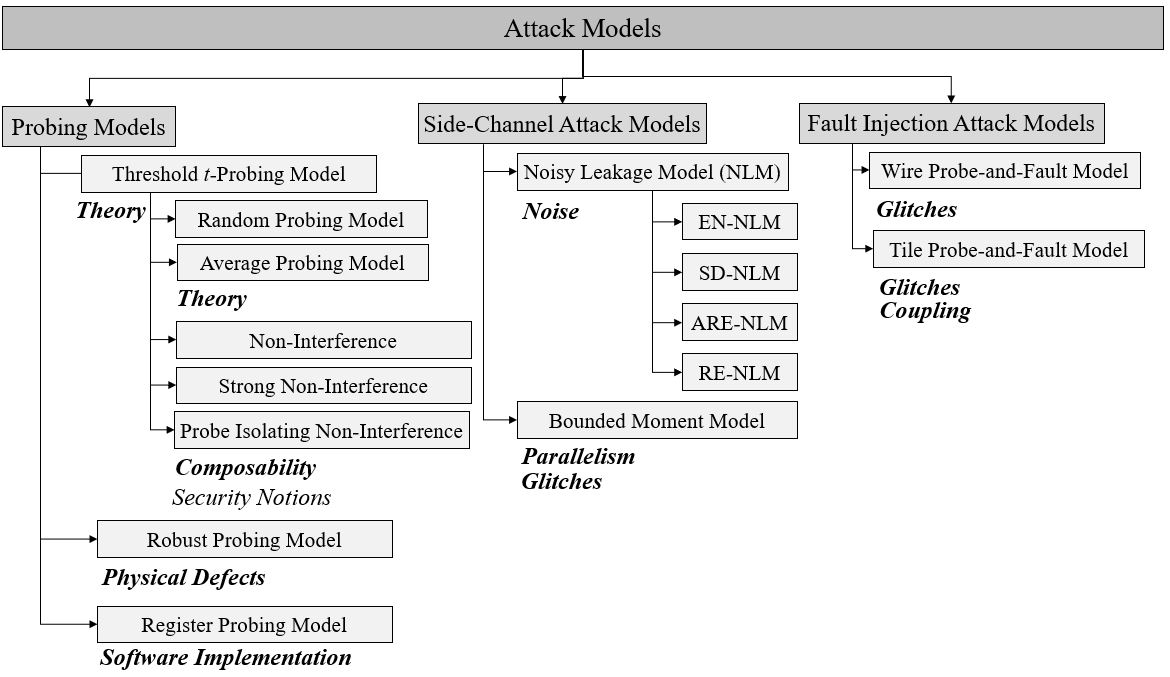}
  \caption{A taxonomy for circuit masking schemes, which also serves as organization of this paper.}
  \label{fig:models}
\end{figure}

This section will describe the attack models, as outlined in the taxonomy presented in Figure~\ref{fig:models}. Early attack models have considered probing as a theoretical, worst-case scenario~\cite{chari1999towards}, while the side-channel attacks were considered as realistic attack models.

\subsection{Probing Model}\label{sec:attacks_probing_model}
The probing model in~\cite{ishai2003private} was one of the first used to quantify masking scheme hardness against probing attacks. It considers a scenario in which the adversary can learn information (free from noise) once $t$ wires of the circuit are probed. This is considered a strong attack model that reflects the capability of an adversary by considering the number of probes ($t$) as well as the cost of an attack. 
The adversary is also limited in the ability to observe different wires with the same probe within a single clock cycle. 
Within this context, security proofs are devised with regard to the notion of ``simulation\footnote{It must be emphasized that simulators and simulations in this context are not related to electronic or digital circuit simulation. Rather, the notion of simulation is taken from cryptography and used as \textit{theoretical proof} of desired circuit behavior.}.'' 

\noindent\textbf{Simulation-based Proof of Security: }
Similar to various cryptographic systems, the security of a masked scheme can be proved via a simulation paradigm. 
Simulation concerns the comparison of what happens in the ``real world'' to what happens in an ``ideal world,'' where the primitive in question (i.e., the masked circuit) is secure by definition~\cite{lindell2017simulate}. 
For a masked circuit, this means to compare what can be learned by an adversary who really probes $t$ positions in a circuit to what can be learned by an adversary who probes no positions. 
In other words, the masked circuit is secure in the sense of simulation if the only information extracted (or output by the adversary) is based on a prior knowledge cf.~\cite{lindell2017simulate}. 
To prove this through simulation, one constructs a simulator that resides in the ideal world being secure by definition and fulfills the following.

\begin{enumerate}
    \item The simulator generates the real adversary's view that is computationally indistinguishable from its real view. 
    \item It extracts the effective inputs given by the adversary during the execution. 
    \item The view generated with regard to the inputs must be consistent with the output in the real world.
\end{enumerate} 
For instance, for a gadget, this has been well-formulated in~\cite{dhooghe2020let}. 
The adversary (also called ``distinguisher'') interacts with the real gadget in the real world, and a simulator mimics its interaction. 
The inputs given to the simulator are a few of the input shares, similar to the inputs in the real world. 
The adversary must not be able to determine whether it is interacting with the simulator or with the actual gadget. 
The failure to distinguish between the simulator and the gadget implies that the adversary can know only the shares given to the simulator. 

\noindent\textbf{Security Proofs of Masked Scheme in $t$-probing Model:} 
Following the simulation-based approach, to prove the security in the $t$-probing sense, the view of the adversary without knowing the input values is simulated. 
A gadget is secure in the $t$-probing model if the distribution of the simulated $t$ probed values is \emph{identical} to the distribution that the adversary would obtain in the real world.
In other words, a gadget is secure in the $t$-probing model if the $t$ values probed by the adversary are independent of the sensitive variables. 
In this regard, two probing models were presented in~\cite{ishai2003private}: \textit{threshold-$t$ probing model} and the \textit{random threshold probing model}. The random probing model introduces the average case of security, compared to the worst-case in $t$-probing model. 
The difference between threshold $t$-probing model and random threshold probing model is that to perform the simulation, in threshold $t$-probing model, the simulator's inputs
are chosen uniformly and independently.  
On the contrary, in random threshold probing model, the inputs are chosen independently from a specific probability distribution. 

\begin{figure}
  \includegraphics[width=\linewidth]{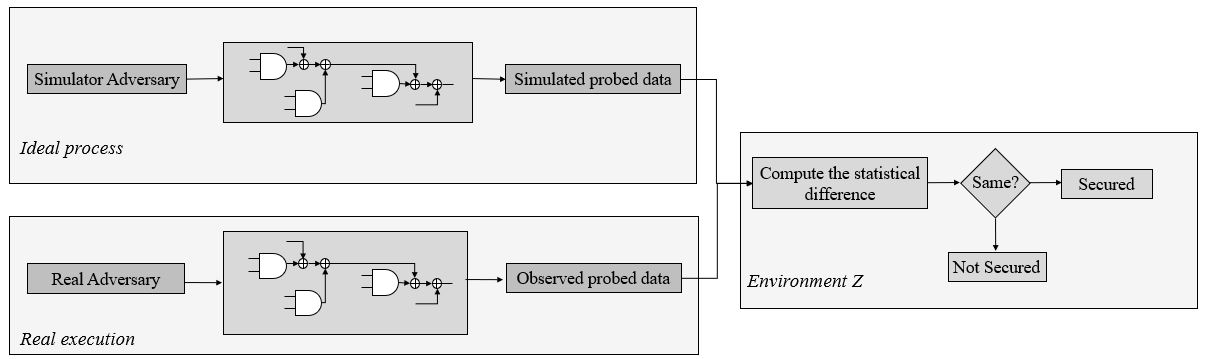}
  \caption{Use of simulator to prove security in probing models}
  \label{fig:simulator}
\end{figure}

\noindent\textbf{Perfectly $t$-probing Secure vs. Statistically $t$-probing Secure:} 
The security proof explained earlier in this section is referred to as perfect $t$-probing security. 
This is due to how the simulation is defined: the distribution of the simulated $t$ probed values is \emph{identical} to the distribution that the adversary would obtain in the real world.
To further evaluate the security of the masked circuit, one can consider the statistical $t$-probing security. 
For this, instead of requiring the distributions to be identical, the statistical distance between the joint distributions of simulated probed values and the joint distributions of probed values is considered. 
Statistical distance determines if the simulator is able to simulate the real probed values given the input shares. 
This process is visually described in Figure~\ref{fig:simulator}.

\noindent\textbf{Private Circuits vs. $t$-probing Secure Circuits: }
Although the notion of privacy and $t$-probing security could be used interchangeably in the literature, there is a subtle difference difference between them cf.~\cite{barthe2016strong}. 
Ishai et al.~\cite{ishai2003private} have defined privacy of a gadget $G$ based on the simulation of any $t$ wires in the circuit. 
On the other, $t$-probing security of a gadget implies that the simulation of $t$ of its intermediate wires\footnote{Intermediate wires carry the intermediate variables during execution of the algorithm represented by the circuit} can be performed using at most $t$ shares of each of its inputs~\cite{carlet2015algebraic,barthe2016strong}.  

\noindent\textbf{Take-home Message: }
The $t$-probing models explained above are powerful tools to assess the security of gadgets, although they suffer from shortcomings reported in the literature. 
The main drawback of the probing models is that 
they do not consider physical effects, such as glitches, transitions, and couplings~\cite{knichel2020silver}. Glitches cancel the desired effect of randomness making input splitting into shares and other randomnesses in the gadget useless. Therefore, other probing models developed for higher-order attacks, when $t > 2$ consider physical effects such as glitches,~\cite{nikova2008secure,fischer2005masking}, as well as security notions of non-interference, strong non-interference~\cite{barthe2016strong}, and probe isolating strong non-interference,~\cite{cassiers2019towards,cassiers2020trivially} which consider composability of gadgets in an  entire circuit instead of just the security of the single gadget. These are discussed next.

\subsection{Side-Channel Attack Models}\label{sec:side_channel_attacks}
The original probing models considered physical probing to nets in the circuit, but are not as realistic for describing side-channel attacks. Consequently, four different aspects have been considered and evaluated in more recent work: composability, noise, physical defects, and implementation.  

\subsubsection{Composability:} 
\begin{figure}
  \includegraphics[width=\linewidth]{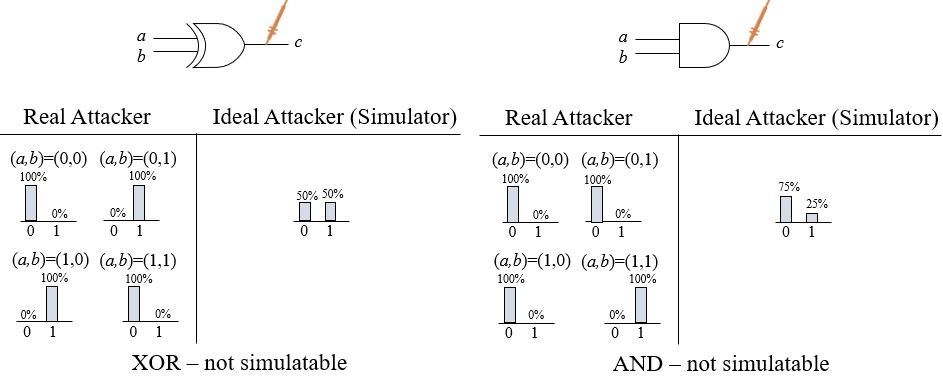}
  \caption{Simulatability definition from~\cite{cassiers2020trivially}: Deterministic Inputs (no inputs given) }
  \label{fig:simulatabilityA}
\end{figure}

\begin{figure}
  \includegraphics[width=\linewidth]{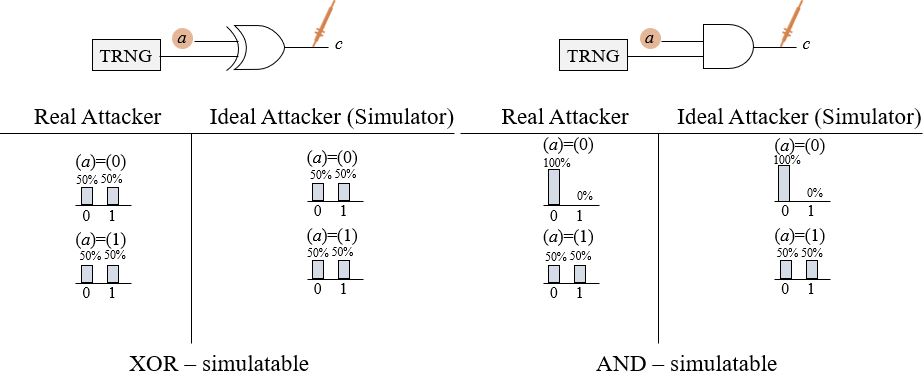}
  \caption{Simulatability definition from~\cite{cassiers2020trivially}: Stochastic Inputs (\textit{a} given) }
  \label{fig:simulatabilityE}
\end{figure}

\textit{Strong Non-Interference and Non-Interference} (SNI and NI) are security notions used to support the composability of gadgets, as the gadgets secured in a probing model do not necessarily produce provably $t$-probing secure circuits~\cite{barthe2016strong}. To define these security notions, one can use the simulatability framework proposed by Bela\"d in~\cite{belaid2016randomness} and illustrated by Cassier in~\cite{cassiers2020trivially}, as shown in Figures ~\ref{fig:simulatabilityA}--, ~\ref{fig:simulatabilityE}. 
These have been usually analyzed through the concept of simulatability defined below. 

\begin{figure}
  \includegraphics[width=\linewidth]{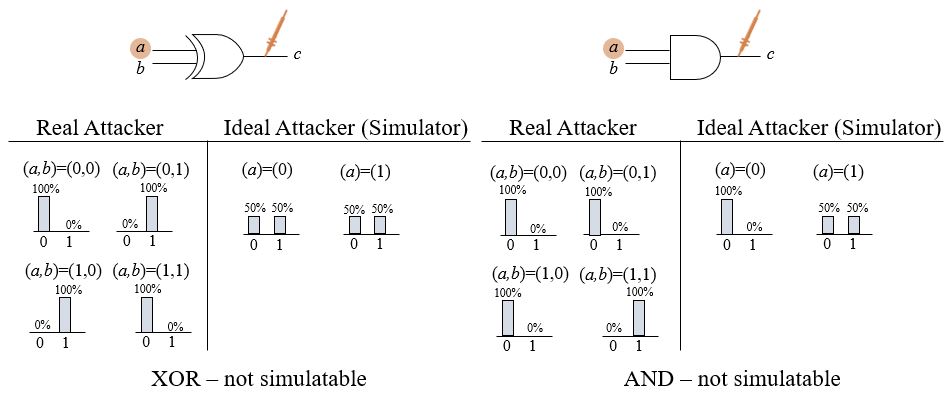}
  \caption{Simulatability definition from~\cite{cassiers2020trivially}: Deterministic Inputs (\textit{a} given) }
  \label{fig:simulatabilityB}
\end{figure}

\begin{figure}
  \includegraphics[width=\linewidth]{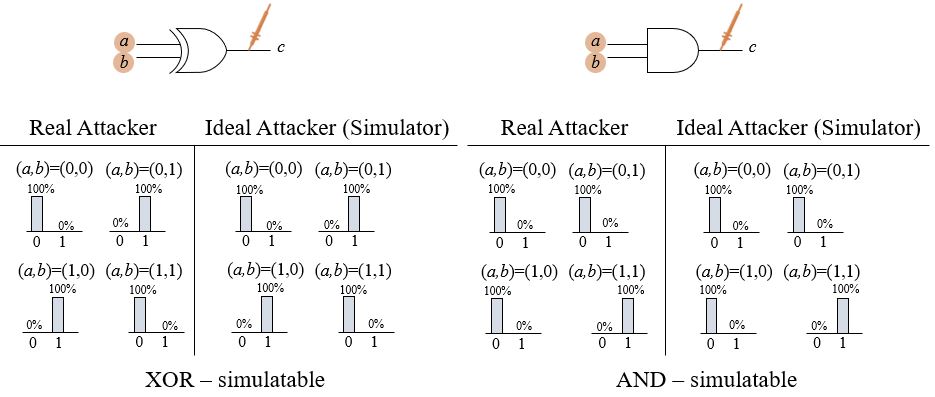}
  \caption{Simulatability definition from~\cite{cassiers2020trivially}: Deterministic Inputs (both inputs given) }
  \label{fig:simulatabilityC}
\end{figure}

\begin{figure}
  \includegraphics[width=\linewidth]{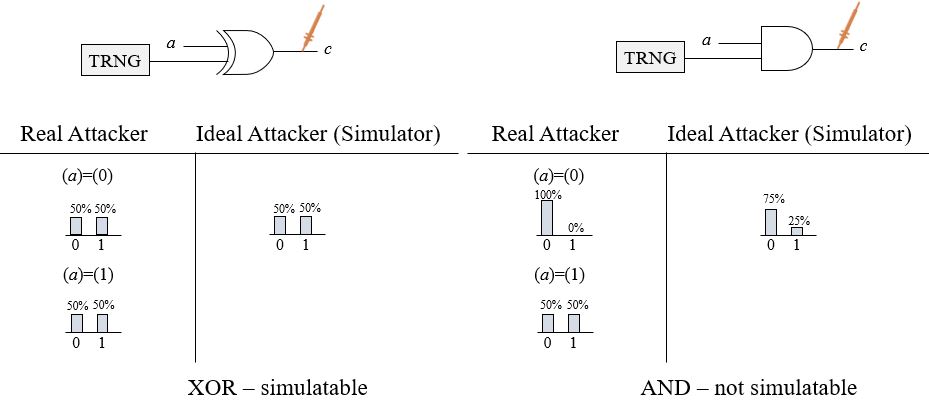}
  \caption{Simulatability definition from~\cite{cassiers2020trivially}: Stochastic Inputs (no inputs given) }
  \label{fig:simulatabilityD}
\end{figure}

\noindent\textbf{Simulatability:} 
According to the simulatability definition, a set of probes can be simulated with at most $t$ shares of each of two inputs if the following holds\footnote{Note that here, simulatability is defined for a gadget with two inputs, although it can be extended to gadgets with more inputs. }. 
A simulator is fed with a set of $2t$ bits output distributions that are identical to the statistical distributions of the probed values in the real world~\cite{belaid2016randomness} (see Section~\ref{sec:attacks_probing_model} for the definition of a simulator). 
In other words, simulatability implies that the true probes are independent of the input shares that are not given to the simulator~\cite{cassiers2020trivially}. 
This independence of the probes from the other shares suggests the $t$ probing security of the gadget. 
Figures~\ref{fig:simulatabilityA}--\ref{fig:simulatabilityE} illustrate the simulatability definition applied on AND and XOR gate for cases where the attacker does not have access to inputs (Figure~\ref{fig:simulatabilityA}), has access to one input (Figure~\ref{fig:simulatabilityB}) and all inputs (Figure~\ref{fig:simulatabilityC}) without randomness, does not have access to any inputs where one input is a random number (Figure~\ref{fig:simulatabilityD}), and has access to one deterministic input and the second input is a random number (Figure~\ref{fig:simulatabilityE}). In all these cases, an attacker probes the output of the AND and XOR gates, while the input is either known or unknown. 
For the sake of simplicity, the inputs are assumed to have one share. 
\begin{itemize}
\item For the case when the attacker does not know the deterministic inputs (Figure~\ref{fig:simulatabilityA}), the distributions between 1s and 0s at the output for all input combinations do not match, making the gates not simulatable.

\item If one input is given, the gates are still not simulatable (Figure~\ref{fig:simulatabilityB}) because the output distribution for the simulator in the ideal world does not match what the real attacker observes at the output. 

\item If both inputs are given to the attacker, the output distributions are the same, making both gates simulatable (Figure~\ref{fig:simulatabilityC}); however, the gadget is not secure due to the lack of enough shares considered for the inputs. 

\item When a random number is introduced as an input of the gate, the XOR is simulatable, whereas the AND gate is not simulatable (Figure~\ref{fig:simulatabilityD}).  
On the other hand, both gates are simulatable if the deterministic input is known to the attacker (Figure~\ref{fig:simulatabilityE}). Nevertheless, the $t$ probing security ($t=1$ in these figures) condition is not fulfilled as the number of shares for each input is not large enough. 
Suppose that each input has $d=t+1=2$ shares (i.e., the best situation studied in the literature~\cite{cassiers2020trivially}), these gadgets are $t=1$ probing secure. 
\end{itemize}

\begin{figure}
  \includegraphics[width=\linewidth]{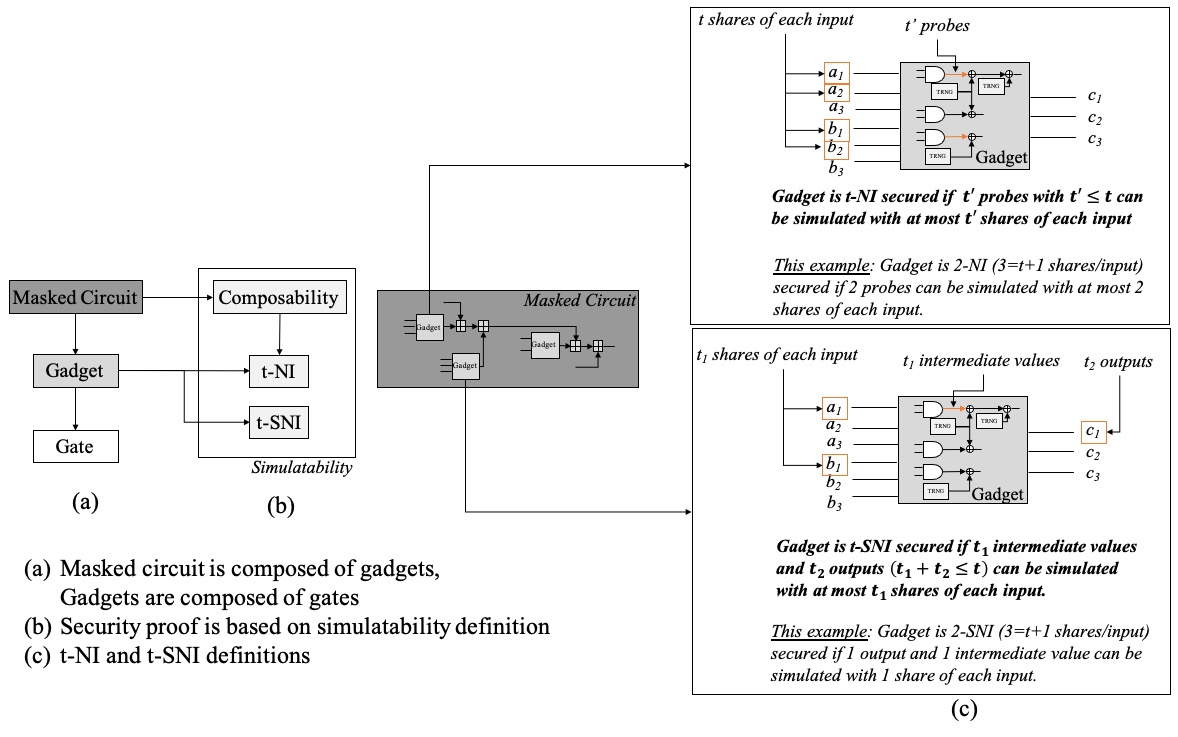}
  \caption{ t-NI and t-SNI definitions}
  \label{fig:sni/ni}
\end{figure}

\noindent\textbf{Security of Composed Circuits:} Composability is considered on the level of a masked circuit, which is a network of connected gadgets. Simulatability is used as a security definition to prove the security through $t$-NI, $t$-SNI, and PINI security notions defined and discussed below. The relationship between these is shown in Figure~\ref{fig:sni/ni}.

A gadget is \textit{$t$-NI secured} if every set of $t' \leq t$ probes can be simulated with at most $t'$ shares of each input~\cite{belaid2016randomness}. 
The notion of $t$-NI security is equivalent to the $t$-probing security as defined in~\cite{carlet2015algebraic}, and perfect $t$-probing security as defined in~\cite{coron2013higher,rivain2010provably}. 
For gadgets with $d >t$ shares ($d-1> t$ to mitigate physical defaults such as glitches~\cite{nikova2011secure}), $t$-NI security is strictly stronger than $t$-probing security. 
In this respect, if a gadget is $t$-NI, then it is $t$-private (as defined by Ishai et al.~\cite{ishai2003private}) secure; however, a $t$-probing secure gadget is not necessarily $t$-NI~\cite{belaid2016randomness,belaid2017private}. 
As well described in~\cite{cassiers2020trivially,barthe2016strong}, 
in the $t$-probing security model (following Carlet et al.'s definition), no statement is given on whether the simulation of the secret shares or any input share should be considered. 
Concretely, to prove the $t$-NI security, the inputs given to the simulator are independent of the shares of the sensitive variables (i.e., the key); hence, the simulated probes are independent of the sensitive input, and the distribution of the true probes are indistinguishable from the distribution of the simulated ones. 
Nevertheless, $t$-NI is not a necessary condition for probing security since the simulated probes for any input shares -- not only for any share of the sensitive input -- must be indistinguishable from the real probes. 

The concept of $t$-NI has been adopted to create composed circuits that remain $t$-NI secure, although in most cases, at a cost of using many refresh gadgets providing the randomness~\cite{goudarzi2017fast}. 
The definition has therefore been extended to $t$-SNI security. 
In fact, the reasoning about the security of composed gadgets has been reinforced by the introduction of $t$-SNI security. 
Intuitively, the $t$-SNI security enables the designer to compose any set of SNI gadgets as it stops the propagation of dependencies from the outputs to the shares of sensitive inputs. 
Specifically, $t$-SNI condition requires that the probes on the output wires do not depend, i.e., (1) if there is no internal probes, their distribution is uniform, (2) otherwise, their joint distribution is determined by only the probes placed on the internal wires. 
More precisely, for a gadget to be \textit{$t$-SNI secured}, $t_1$ and $t_{2}$ probes at the input shares and output of the gadget with $t_{1} + t_{2} \leq t$ can be simulated, knowing $t_{1}$ shares of each input.  
For example, in the Figure~\ref{fig:sni/ni}, the gadget is 3-NI secured if one probe at the output, and one probe inside of the gadget can be simulated knowing 2 shares of each input. The same gadget is 3-SNI secured if those probes can be simulated with only one share of each input.


\begin{figure}
  \includegraphics[width=\linewidth]{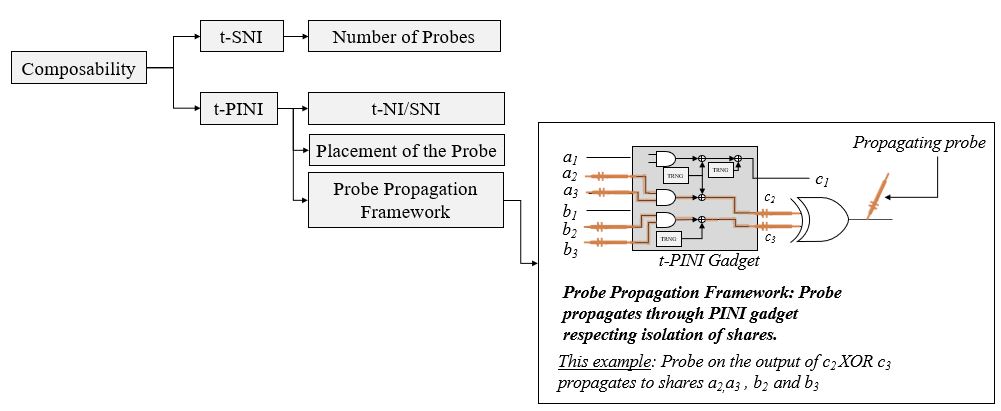}
  \caption{ Circuit including PINI AND Gadget inspired by~\cite{cassiers2020trivially}. Orange probe is the probed output, and dashed orange wires are the \textit{propagated} probes, which may or may not propagate according to the simulatability definition. If the probe can propagate freely towards the inputs, those input values can be learned}.
  \label{fig:PINI}
\end{figure}

The SNI secured circuit requires a large number of random number generators because a linear function\footnote{A linear Boolean function $f$ over $\mathbb{F}_2$ is defined as $f(c)=f(c_1, c_2, \ldots, c_n)=\sum_{i \in S} c_i.$ for some set $S\subseteq \{1, \ldots, n\}$. For instance, XOR is a linear function over $\mathbb{F}_2$. } is not trivially SNI secured (i.e., trivial implementation means that functions can be implemented share by share). Therefore,~\cite{cassiers2020trivially} introduced the \textit{Probe Propagation Framework and Probe Isolating Non-Interference} (PINI), which extends the $t$-S/NI definitions. In PINI, if the attacker probes the output, input values can be simulated in a backward fashion, as shown in Figure~\ref{fig:PINI}.  
In this figure, if the attacker probes the wire where the probe is, the probe \textit{propagates} back in an isolated fashion. In other words, the output probe can only learn information from input shares that are in its fanin. If shares of the sensitive input can be learned, the circuit is not secured. PINI security requires that all gadgets must be $t$-NI. According to the PINI security notion, more than $t$ shares cannot be learned, and the total number of probes placed on intermediate values and outputs must be less or equal to $t$ in SNI gadgets. The PINI security notion depends on the placement of the probe as well as the number of probes.

Formal proofs of probing, non-interfering, and strong non-interfering security notions are described in~\cite{belaid2018tight}, in which these three definitions are further analyzed and proved based on \textit{concrete security games}. In those games, the attacker interacts with simulator learning circuit values in steps based on simulatability definitions.    

\subsubsection{Noise:} 
Side-channel leakage is often noisy. For example, power traces contain algorithmic noise from computations that are not correlated with the secret under attack~\cite{standaert2013leakage} and measurement noise such as quantization error. Thus, to more realistically represent a side-channel attacker, the \textbf{noisy leakage model} was introduced in~\cite{prouff2013masking}. In this model, the adversary has access to the input $x$ and the leaked noisy data $f(x)$, where $f(\cdot)$ is a randomized function that captures the impact of the noise through the distance between the distribution of $x$ denoted by $\Pr(x)$ and the conditional distribution $\Pr(x|f(x)$, so-called bias. 
The main assumptions are that the noise level is chosen by the designer and linearly bounded, and the statistical distance between the distributions $\Pr(x)$ and $\Pr(x|f(x))$ is bounded by $\delta$. 
In~\cite{prouff2013masking}, this distance is measured by using the Euclidean Norm (EN). The proposed model is very detailed, depicting every processing step and any type of noise, as long as it satisfies statistical metrics. 
Nevertheless, this noisy leakage model is proved using an information-theoretic security proof which brings its limitations~\cite{kalai2019survey}. The first limitation is that it considers a true random number generator(TRNG), which is tamper-proof and does not leak any information\footnote{A leak-free refresh operation takes as input an encoding $(X_1,..., X_d )$ of a secret $s$ (i.e., shares of the secret) and outputs a freshly and independently chosen encoding (i.e., new shares) of the same value~\cite{duc2014unifying}.}
The second limitation is the consideration of random message attacks, which assumes that secret inputs are uniform random values and does not consider oracle (input/output) attacks. The final limitation is that noise must be sufficiently large, which often fails to hold in practice.

Information-theoretic security in~\cite{prouff2013masking} gives a bound for mutual information between the key and leakage from cipher without considering the joint distribution between the leakage and input-output pairs. 
However, this noisy leakage model is very complicated, and it is not easy to use for state-of-the-art masking schemes according to~\cite{dziembowski2008leakage}. Therefore, other leakage models applying different statistical noise metrics have been developed that could better reflect the specific properties of circuit masking schemes. The first example is the use of statistical distance (SD), instead of Euclidean Norm, in~\cite{duc2014unifying,duc2019making}. The same noise characteristics are considered, where inputs $x$ are biased rather than uniform, and the noise function is $\delta$-noisy following the concept of indistinguishably of distributions. 
The security of EN- and SD-noisy leakage models has been reduced to the security of random probing model from~\cite{ishai2003private}, and average probing model~\cite{dziembowski2015noisy}. The average probing model is similar to the random probing model, although the noise leakage probability $\epsilon$ is is taken over the input $x$. 

The two newest noise leakage models are Relative Error (RE) and Average Relative Error (ARE)~\cite{prest2019unifying}, which show tighter relations to the $t$-probing model and random probing model,~\cite{ishai2003private}. For these new models, the probes are simulated, and the simulator outputs exactly the same distributions despite different metrics because the distributions depend on intermediate values, as discussed for $t$-probing model. 
Nonetheless, RE-noisy leakage model, similar to the Average random probing model, and the EN-noisy leakage model, considers leak-free refresh gadgets, which is its main drawback. 
This issue is resolved in ARE-noisy leakage model. 

\subsubsection{Implementation:} 
It has been reported that the probing model better captures the characteristics of serial implementations, assumed when analyzing the security of software implementation~\cite{reparaz2015consolidating}. 
Although parallel implementation is possible for modern microprocessors, this type of implementation has been mainly practiced for hardware platforms. 
According to~\cite{barthe2017parallel}, the noisy leakage model in~\cite{prouff2013masking} can inherently be used for implemented parallel circuits; however, their security is harder to prove in the usual formal models for the verification of cryptographic implementations. 
This is due to the fact that the security proofs based on noisy leakage model require manipulation of leakage distributions in the formal models. 

To address this, Barthe et al. introduce the \textbf{bounded moment model}~\cite{barthe2017parallel}, which mitigates the statistical recombination of inputs' shares. The motivation behind this model is that the security of a masked circuit against side-channel attacks comes from the requirement of estimating higher-order statistical moments, which gets exponentially harder for a greater number of probes, assuming sufficient noise.  
Another probing model is the \textbf{register probing model},~\cite{belaid2020tornado} used for masked software implementations. The implementation works on registers that hold multiple bits, which leak all \textit{together} as compared to $t$-probing model where one probe extracts one bit. The security definitions are based on concrete security games, as in~\cite{belaid2018tight}, with the main difference that the attacker now learns a vector of secret values. 

\subsubsection{Physical defaults:} Robustness against physical defaults, such as glitches (compromising randomness provided by refresh gadget)~\cite{prouff2011higher}, recombination of defects, and composition have been considered in the \textbf{robust probing model} by~\cite{faust2018composable,molteni2020spectral}. Its glitch-resistance comes from composability (SNI property), and \textit{non-completeness property}, which is used to prevent combinatorial logic from having access to all shares of secret input in its fanin~\cite{faust2018composable}.

\subsection{Fault-Injection Attack Models}
\begin{figure}
  \includegraphics[width=0.65\linewidth]{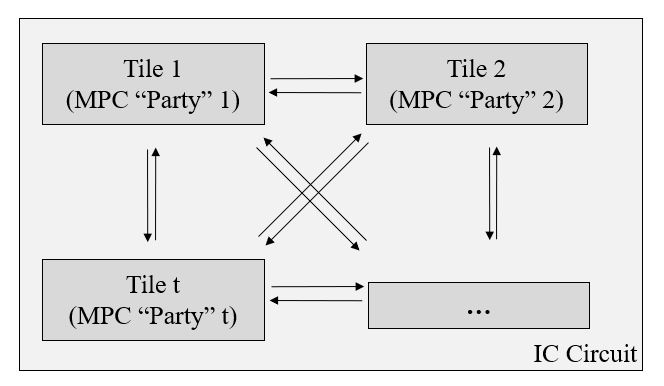}
  \caption{Tile-Probe-and-Fault Model partitions circuit into Tiles, where each Tile represents one party in communication protocol}.
  \label{fig:mpctile}
\end{figure}

Active attack models have been discussed in~\cite{reparaz2018capa,dobraunig2020exploring}.  \cite{reparaz2018capa} has introduced the \textbf{tile-probe-and-fault model}, which is an extension to \textbf{wire-probe model}~\cite{ishai2006private}. Fault attack models consider the ability of an attacker to inject faults into a circuit in terms of the type of the fault, location, and duration. Tile-probe-and-fault model partitions the circuit into blocks called ``tiles,'' which mimics a multi-party computation (MPC) protocol, as visually shown in Figure~\ref{fig:mpctile}. In this model, it is assumed that the leakage of each tile is independent of others, which means if the attacker observes leakage of one tile, no information can be extracted from other tiles in the circuit. In this model, an attacker extracts information through side-channel analysis and by injecting faults into the circuit. The goal of this model is to detect the fault as opposed to correcting it~\cite{reparaz2018capa}. 

\begin{figure}
  \includegraphics[width=0.65\linewidth]{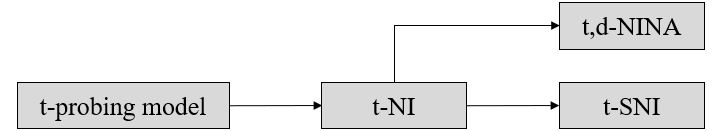}
  \caption{Non-Interference (NI) and Non-Accumulation (NA) NINA security notion has been developed from the $t$-NI security notion, inspired by~\cite{dhooghe2020my}}.
  \label{fig:nina}
\end{figure}

\begin{figure}
  \includegraphics[width=0.5\linewidth]{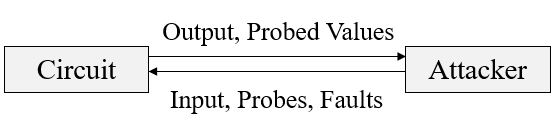}
  \caption{The communication between attacker and circuit. Attacker introduces the faults into circuit, where fault does not accumulate and only affects one share, inspired by~\cite{dhooghe2020my}}.
  \label{fig:ninanesto}
\end{figure}

The model of Non-Interference and Non-Accumulation (NINA) has been introduced in~\cite{dhooghe2020my}, where NINA is closely related to $t$-NI security notion, as shown in Figure~\ref{fig:nina}. In this model, an attacker introduces the fault into a circuit, but it does not accumulate: only affecting the one input share that can be learned using error-correcting codes. The communication between circuit and the attacker is shown in Figure~\ref{fig:ninanesto}. Independent NINA notion relaxes NINA in terms of separating the effect of faults and probes, without requiring one to feed the security verification simulator with shares for every fault in the circuit. These notions have been introduced on two gadgets: error detection and error correction~\cite{dhooghe2020my}.

The attack models explained in this section are used for development of masking schemes with provable security, specifically AND gadgets. Section~\ref{sec:masking_schemes} will describe the gadgets used in implementable schemes in Section~\ref{sec:masking_implementations}.


\section{Masking Schemes}
\label{sec:masking_schemes}
From the perspective of cryptography, masking is an XOR-secret sharing developed as a countermeasure against leakage from computation~\cite{kalai2019survey}. 
When first being put forward, masking concerned how to keep individual wire values in a design, e.g., smart-card circuit, independent of the secret key~\cite{chari1999towards,goubin1999and}. 
In this regard, each wire carrying a bit $b$ is usually replaced by a set of wires carrying the Boolean masked of $b$ so that the exclusive-or of the values carried by the set of wires is equal to $b$. 
Soon after its introduction, masking became one of the topics that have attracted a great deal of attention from researchers working on not only the theory of leakage-resilient cryptography, but also the secure design of primitives. 
This section explores some ingredients of masking, including gadgets and randomness, as well as types of masking. 

\subsection{Gadgets}


As introduced earlier, gadgets are the main building blocks of any masking scheme. Masking schemes consider each gate of the original leaky circuit to be transformed into a leak-free cluster of gates while preserving the original functionality. Linear gates\footnote{The gate function in $\mathbb{R}_2$ can be defined from the truth table, where the OR represents summation ($+$) and AND represents multiplication ($\cdot$)}, such as NOT, XOR and XNOR, have not been of interest to the research because of their commutative and associative properties, similar to addition. On the other hand, non-linear AND gate has been extensively researched. During the early development of attack models, as explained in Section~\ref{sec:attack_models}, various gadgets have been developed to prove the security of attack models~\cite{canright2009very,ishai2003private,prouff2013masking,schramm2006higher,coron2013higher,belaid2017private,coron2014higher}. However, those usually do not yield securely implemented masking schemes. 
For that purpose, although the security of most attack models is reduced to the security of a gadget, masking implementations should focus on the \textit{composition} of gadgets and the \textit{use of randomness}. 

The main goal of composing the gadgets remains the same: the functionality of the composed gadget resembles the functionality of the gate, e.g., the AND gate. If we consider an AND gate with original inputs $a$ and $b$, to provide security in the sense of $t$-probing model, those inputs are shared (or split) into $2t+1$ secret shares where $a=a_1\oplus{a_2}\oplus\cdots\oplus{a_{2t+1}}$, $b=b_1\oplus{b_2}\oplus\cdots\oplus{b_{2t+1}}$. Shared value ($a$, $b$) is independent of its shares, which are randomly drawn from a uniform distribution. In practice, this is accomplished by using a random number generator (RNG), or the so-called \textit{refresh gadget}~\cite{goudarzi2018secure}. Refresh gadget is usually considered as a leak-free component, and therefore, recognized as a drawback in various attack models, as already discussed in Section~\ref{sec:attack_models}. Computation is performed on shared input values, which are masked by random numbers computed from the refresh gadget.

This subsection introduces the main building blocks used in implementation schemes discussed in Section~\ref{sec:masking_implementations}.

\subsubsection{ISW AND gadget:} 
\begin{figure}
  \includegraphics[width=0.45\linewidth]{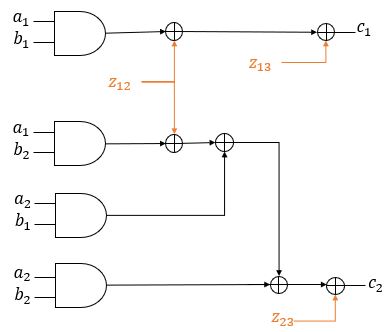}
  \caption{ISW AND Gadget}
  \label{fig:ISW_AND}
\end{figure}

The most important construction is from~\cite{ishai2003private}, with its so-called \textit{ISW AND gadget}. ISW AND gadget has been proven secure in the threshold $t$ model and in various noisy leakage models. ISW AND gadget expands one AND gate into $t^2$ gates, where values need to be split into $2t+1$ shares to be secured against $t$ number of probes. The reason for $2t+1$ instead of $t+1$ shares in ISW AND gate is due to the computation sequence, shown in Figure~\ref{fig:ISW_AND}, which has to be preserved to claim the security at $t$ security order. Therefore, the order of handling intermediate values plays an important role in the definition of security for the specific gadget, on the gate level. 

ISW AND gadget with shared inputs $a$ and $b$, and shared output $c$, is constructed as follows: \begin{enumerate}
\item Random $z_{i,j}$ values are drawn from refresh gate uniformly for $1\leq i\leq j\leq 2t+1$.
\item Intermediate values $z_{j,i}$ are computed using $z_{j,i}=(z_{i,j}\oplus{a_ib_j})\oplus{a_jb_i}$, where parentheses signify the order of computation.
\item Finally, shared output values are computed using $c_i=a_ib_i\oplus{\bigoplus_{i\neq{j}}{z_{i,j}}}$.
\end{enumerate} 
An ISW AND gadget shown in Figure~\ref{fig:ISW_AND} is secured against 1 probe (i.e., $t=1$), with three shares per input ($2t+1=3$), three shares per output $c$ and uniform, and random masks $z_{12}, z_{13},z_{23}$. 
This gadget can be extended to higher order protection schemes, by building the scheme based on multiplication computation, instead of AND gates~\cite{reparaz2015consolidating}, which is very useful in implementations of AES. ISW AND gate also is used as the basis for \textit{threshold implementations} explained in the next section. 

\subsubsection{Trichina AND Gate:} Another important gadget used for \textit{consolidating masking schemes (CMS)} is the Trichina AND gate, introduced by Trichina in~\cite{trichina2003combinational}. Trichina gate is meant only to be secured against $t=1$ on the gate level, while ISW gate can be applied to the circuit level considering the composition of the gadgets. 
Trichina AND gate requires two 2-share inputs (e.g., $a_1$ and $a_2$ for input $a$) and 1 bit $z$ of randomness. The order of operations in Trichina AND gate is very important, as it is the basis for security. Trichina AND gate produces two outputs for a 2-input (4 shares) AND gate, as shown in Equation~\ref{equation_trichina}~\cite{reparaz2015consolidating}:
\begin{align}
\label{equation_trichina} 
	&c_1 = \left((z\oplus{a_1b_1})\oplus{a_2b_1}\right)\oplus{a_1b_1} \nonumber\\
	&c_2 = z.
\end{align}

\subsection{Randomness}

Randomness is a critical element in any masking scheme's development and has been the subject of ongoing research. Creating random numbers at high throughput in hardware is a complex and costly process that imposes high energy consumption and area overhead. Randomness in masking schemes have been analyzed from different angles:
\begin{itemize}
    \item Designing the most optimized refresh gadget
    \item Optimizing randomness of the entire circuit (i.e., masking scheme)     
    \item Nullifying \textit{online} randomness and considering only initial randomization of inputs (i.e., input sharing)
\end{itemize} 

These topics are discussed in this section and the definition of usual terms is provided in Table~\ref{tab:terminology_randomness}. 


\begin{table}[t]
\caption{Terminology Related to the Randomness in Masking Schemes}\label{tab:terminology_randomness}
\centering
\small
\begin{tabular}{ll}
\cline{1-2}
\multicolumn{1}{l|}{\textbf{Term}} & \multicolumn{1}{l}{\textbf{Definition}}  \\ \cline{1-2}
\multicolumn{1}{l|}{Refresh Gadget} & \multicolumn{1}{l}{Algorithm which introduces random number} \\ \cline{1-2}
\multicolumn{1}{l|}{Online Randomness} & \multicolumn{1}{l}{\begin{tabular}[c]{@{}l@{}}Introducing random numbers to intermediate values of the gadget\\ with help of refresh gadget\end{tabular}} \\ \cline{1-2}
\multicolumn{1}{l|}{Input Randomness} & \multicolumn{1}{l}{Input value being split into shares with help of refresh gadget} \\ \cline{1-2}
\end{tabular}
\end{table}

\subsubsection{Refresh Gadgets} Refresh gadgets, which are also called \textit{refreshing algorithms}, are an influential factor in controlling the area overheads in masking scheme implementations~\cite{barthe2020improved}. 
Therefore, the design of refresh gadget with a fewer number of random bits has been sought and discussed in the literature. 
For instance, the parallel refresh gadget, from~\cite{barthe2017parallel}, has features which make it useful for hardware implementations, such as~\cite{gross2017reconciling}. 
It is achieved with a simple rotation and XOR operations. Inputs are first loaded, then rotated along with the loading of refresh values. Finally, refresh values are XOR-ed with rotated values. Note that the security of this refresh gadget is proved in the bounded moment model for $d\leq 7$, when an attacker can observe the leakage continuously. 
In fact, in this case, the proposed refresh gadget is not secure in the probing model. 
Another remark is that the proposed refresh gadget can be used to construct an SNI mask refresh gadget fulfilling the properties of composability. 


As a follow-up to the above-mentioned study on refresh gadgets, Barthe et al. have introduced parameterized non-interference ($f-NI$) to prove the security of their gadget for arbitrary orders~\cite{barthe2020improved}. This security notion subsumes SNI and NI by realizing the difference between input/output vs. internal observations. 
To further optimize the implementation of the refresh gadget, it is suggested to rely on an observation made about the pairs of random variables. 
For instance, it is observed that refresh gadgets at low orders rarely use repeated pairs of random variables ($a_i \oplus r_j \oplus r_k$ and $a_j \oplus r_ j \oplus r_k$ with  $i \neq j$ ). 
Based on this, the refresh gadget is updated by changing the way the random values are rotated. 
Another optimization comes in terms of decreasing the leakage of intermediate values by changing the place of random value insertion; instead of immediate insertion, a random value can be inserted into an encoding of 0 using the same rotation-based technique, and then it can be inserted into a signal. This makes the random value \textit{locked}, and it does not leak the information possible for attacks~\cite{barthe2020improved}. 

Finally,~\cite{belaid2016randomness} compares the masked multiplication (AND) using two TRNGs: ISW and parallel refresh gadget. Parallel refresh gadget outperforms ISW by 25-40\% in speed, and it requires less randomness than ISW multiplication.

\subsubsection{Online Randomness} This is the randomness used for refreshing the intermediate values in masking schemes. The randomness required for masking implementations has decreased throughout the years. Initial randomness requirement was $(d+1)^2$~\cite{de2016masking}. It was then reduced to $d(d+1)/2$~\cite{gross2017efficient,gross2016domain}, and finally to $\lfloor d^2/4\rfloor +d$~\cite{belaid2016randomness,gross2017reconciling}. 
Various masking schemes have been developed with the goal of decreasing the randomness, such as recycled randomness masking (RRM) and reduced randomness shuffling (RRS)~\cite{papagiannopoulos2018low}, and low entropy masking schemes (LEMS)~\cite{grosso2013low}. However, these masking schemes tailored towards reduction of randomness are risky to implement because hardware assumptions of constant parameters are hard to control and make tamper-proof~\cite{austrin2014impossibility}. 
Random number generators, 64-bit LFSR, have been implemented on FPGA for CMS and domain-oriented masking (DOM) designs in~\cite{de2018hardware}. Compared to TI, CMS and DOM require fresh randomness during the operation of non-linear gates. DOM requires half of the randomness of CMS~\cite{de2018hardware}.

\subsubsection{Input Randomness}
Classical threshold implementations that require $d+1$ shares per input~\cite{reparaz2015consolidating,gross2016domain}, usually require fresh randomness at intermediate values. The work in~\cite{moradi2020reconsolidating} and~\cite{gross2019first} focus on a masking scheme, which does not need any fresh randomness, which means once the input is shared (masked), there is no need to re-mask it inside the circuit for $d=1$ and two input shares. 
The work in~\cite{moradi2020reconsolidating} focuses on 180nm ASIC implementation secure in the glitch-extended probing model. They have implemented this scheme on different designs of s-boxes without any fresh randomness. Compared to the current state-of-the-art, this is the only implementable masking scheme that does not require fresh randomness. However, it is limited to an attacker with $d=1$ probes. According to~\cite{moradi2020reconsolidating}, their algorithm can be extended to higher security orders, but it would not completely nullify online, refreshing randomness. Their results are summarized in Table~2 in~\cite{moradi2020reconsolidating}. 
On the other hand, the study in~\cite{gross2019first} focuses on the implementation of 2-share masking in ARM processor Cortex-M4 with summarized work in Table~7 in~\cite{gross2019first}. This work has emphasized that 2-share masking schemes without online randomness still cannot be easily compared across the field, as many papers do not report the same information.

\subsection{Boolean Masking}


Boolean Masking, with gadgets as a building block, has been applied to various cryptographic algorithms. To the best of our knowledge, an open-source automated tool which creates masked circuits and masked implementation of entire cryptographic algorithms is not available. However, various works have proposed formalization of \textit{compilers}, where the input into the compiler is a circuit description (e.g., netlist), and the output is a protected circuit. The formalization provides insight into how the original circuit has to be masked, such as in~\cite{andrychowicz2016circuit,moss2012compiler}, which describes the amount of leakage from the transformed circuit. According to~\cite{eldib2014synthesis}, side-channel attack masking countermeasures have previously been synthesized using known code patterns. However, it focuses on creating an SMT solver-based exhaustive search for \textit{perfectly} masked circuit called \textit{SC Masker}. 

The research has focused on defining the framework, such as \textit{maskComp} from~\cite{barthe2016strong} used for SNI/NI notions. Algorithms such as this one produce small proof-of-concept circuits, which are further validated and verified, usually in the same work. Verification tools, which will be introduced in Section~\ref{sec:masking_implementations} are developed and used to verify the security of reported and publicly available masked circuits.    

\subsection{Beyond Boolean Masking}
Various masking techniques have been developed to either augment or replace Boolean masking, as reported in Table~\ref{tab:BeyondB}.

\begin{center}
\small
\begin{longtable}[t]{p{3cm}|p{4cm}|p{2.5cm}|p{2.5cm}}
\caption{Beyond Boolean Masking Schemes}
\label{tab:BeyondB}\\\hline
\textbf{Masking} & \textbf{Compared to Boolean Masking} & \textbf{Fault Injection Attacks} &\textbf{Side-Channel Attacks}\\
\hline
Polynomial Masking & Greater algebraic complexity & & \checkmark \\
\hline
Shamir's Secret Sharing Masking & Same security, better efficiency &  & \\
\hline
Inner Product Masking & Better security, larger area and complexity & & \\
\hline
Direct Sum Masking & Higher complexity & \checkmark & \\
\hline
Code-Based Masking & Higher complexity & \checkmark & \\
\hline
Multiplicative Masking & Augmentation to Boolean Masking of AES & & Vulnerable against DPA \\
\hline
Homographic Functions Based Masking & Focused on $d>2$ attacks & & \\
\hline
Affine Masking & Focused on $d>2$ attacks, Security and performance trade-off & & \\
\hline
Coding Theory Based Masking & Focused on $d>2$ attacks & & \\
\hline
\end{longtable}
\end{center}

\subsubsection{Polynomial Masking} Polynomial masking (PM), introduced in~\cite{prouff2011higher} combines Shamir's sharing scheme and multi-party computation techniques according to~\cite{carlet2019polynomial} with a goal of protecting against physical glitches. In PM, the basic masking computation is done by function $\Sigma{(ZX)}$, where $Z$ is a mask vector, $X$ is a secret vector, and $\Sigma$ is XOR operation. The purpose of the mask vector $Z$ is to hide the real values of secret vector $X$, where the output is $Z\oplus{X}$. Polynomial masking scheme has been compared to Boolean masking by studying the physical leakage of $d$-tuple of shares $S$ with $d+1$-tuple of leakages of $HW(X)+\beta$, where $HW(\cdot)$ is the Hamming weight of secret shares, and $\beta$ is Gaussian noise with mean 0 and standard deviation $\sigma$. This scheme has been applied on a bit-sliced paralleled AES circuit in~\cite{goudarzi2017fast} showing the \textit{superiority} of bit-slicing techniques in regard to circuit masking.

\subsubsection{SSS Masking}
One of the alternatives to Boolean masking schemes was introduced in~\cite{goubin2011protecting} based on 1979 Shamir's Secret Sharing Scheme~\cite{shamir1979share}. This masking scheme uses the same multi-party protocol as in~\cite{prouff2011higher} used for early AES implementations. It has been concluded that diversity among the masking schemes can introduce both better security and lower complexity. 

\subsubsection{Inner product masking (IPM)}
Many practical implementations use Boolean masking as it is simple, requires less area, and is easier to implement~\cite{wang2016inner} than any other masking scheme. However, under more scrutiny, the so-called \textit{inner product masking} is considered an even better option than Boolean masking because of its greater security against higher-order attacks, with the trade-off of area overhead caused by high algebraic complexity~\cite{cheng2020optimizing}. This masking scheme was proposed by Balasch et al. in ~\cite{balasch2012theory}, and further improved and extended in~\cite{balasch2015inner} and~\cite{balasch2017consolidating}. 

In IPM, instead of \textit{just} XOR-ing a random mask with an input signal or intermediate value, it first multiplies masks $Z$ by public vector $L$, and then they are added to the secret data $X$, such that computation is equal to $X\oplus{L}$. However, the number of bits in vector $L$, and the number of masks directly affect the security performances. As noted by~\cite{cheng2020optimizing}, for instance, the scheme involving 2-bits mask and 8-bits $L$ results in security in bounded moment model with $d=3$, while in threshold-probing model with only $d=1$. Various security metrics have been developed to study the security of this masking scheme, such as mutual information (MI), success rate (SR), and signal-to-noise ratio (SNR). This masking scheme in its essence, consolidates multiplicative, affine~\cite{fumaroli2010affine}, polynomial and Boolean masking into one masking scheme.

\subsubsection{Direct Sum Masking}
Direct Sum Masking~\cite{carlet2015complementary, bringer2014orthogonal} is a masking scheme that can be analyzed from an error-correcting code viewpoint, as it was designed to also protect against fault-injection attacks.
The work in~\cite{poussier2017connecting} showed that IPM could be seen as a case of \textit{Direct Sum Masking (DSM)} scheme. Compared to IPM, DSM adds an arbitrary amount of random bits. This masking scheme provides more algebraic complexity, where the masking computation is done by $XG \oplus{ZH}$, where $X$ and $Z$ are secret input, and random masks respectively, but $G$ and $H$ represent generator matrices of the vector spaces designated for secret input and masks, respectively. The work in~\cite{carlet2019polynomial} proposed a practical DSM scheme for AES. It detects and corrects errors in a manner that decreases the memory requirements and computation time. 

\subsubsection{Code-based masking}
Code-based masking (CBM)~\cite{wang2019provable,wang2020efficient} provides security in \textit{low-noise conditions} by increasing the statistical security order (i.e., security order in the bounded moment leakage model) based on its high algebraic complexity. It is designed to also protect against fault injection attacks. The work in~\cite{wang2020efficient} introduces the first generic encoder as a generalization of the former code-based masking, which is applicable to any circuit. It utilizes the properties of parallelism to increase the masked circuit's efficiency by amortizing randomness and computation cost. The multiplication (AND) gadget in this work is explained with respect to coding theory terminology. However, the following explanation \textit{translates} the construction to circuit terminology: 
\begin{enumerate}
  \item Input to the AND gadget is a $d$-bit signal, which gets transformed into ISW AND gadget. This gadget is then extended by AND-ing and XOR-ing the output shares with a pre-computed set of values (which correspond to the generator matrix of the encoder). This computation \textit{transforms} Boolean masking into additive masking where XOR-ing corresponds to addition. 
  \item In the second step, these additive sharings are further \textit{transformed} into a \textit{linear codeword} by AND-ing the previous state with another matrix $A$.
  \item Finally, additional random numbers are introduced into the gadget and run through the ISW-like gadget algorithm to create the final result.
\end{enumerate}

For fault injection security, it is assumed that faults can be injected at the input or output of the gadget while the gadget itself is tamper-proof. Otherwise, it would create the issue of \textit{fault propagation}, which requires additional protection mechanisms~\cite{wang2020efficient}.

\subsubsection{Multiplicative Masking Scheme}
Multiplicative masking was introduced in~\cite{trichina2002simplified}, where it was combined with Boolean masking to protect different portions of AES. However, soon after its introduction, it was shown to be vulnerable against first-order differential side-channel attacks~\cite{golic2002multiplicative} because the zero element cannot be masked with a multiplicative masking scheme~\cite{de2018multiplicative}. Eventually, this was overcome in~\cite{genelle2011thwarting}. In multiplicative masking, the original input is equal to the shares AND-ed with each other (i.e., $a=a_1 a_2 \cdots a_{d+1}$) while in Boolean, the original input is equal to the shares XOR-ed among themselves (i.e.,$a=a_1\oplus{a_2}\oplus\cdots\oplus{a_{t+1}}$).

\subsubsection{Other Masking Schemes} Others worth mentioning studies are based on homographic functions~\cite{prouff2010attack}, affine masking~\cite{fumaroli2010affine}, arithmetic masking\cite{bettale2018improved,coron2017high}, and coding theory based masking~\cite{castagnos2013high}.


\section{Masking Implementations}
\label{sec:masking_implementations}


Masking algorithms 
tailored specifically for cryptographic implementation are based on secret sharing, but the differences come in their implementation of logic functions~\cite{bilgin2014more}. These algorithms focus on practicality and realization in terms of area overhead, speed, randomness requirement, and they have usually been applied to cryptographic algorithms, such as Advanced Encryption Standard (AES), and ASCON, a finalist of CAESAR competition for Authenticated Encryption~\cite{Bernstein2014}. The implementation schemes discussed in this survey are Threshold Implementations (TI), Domain Oriented Masking (DOM), Consolidating Masking Schemes (CMS), Unified Masking Approach, and Generic Low-Latency Masking (GLM). These implementations must satisfy requirements of correctness, non-completeness, uniformity, independent secret sharing, SNI, and probing security. However, De Meyer et al. in~\cite{de2019consolidating} consolidate those \textit{necessary} security conditions into \textit{$t$-glitch immunity} which is \textit{sufficient}, in addition to necessary. 
To compare the implementations among each other, AES is the most common benchmark, and the schemes use \textit{gate equivalent} [GE], a unit of measure of area in terms of two-input NAND gate. The comparison is shown in Table~\ref{tab:tableIMPL}.

\subsection{Implementation Schemes}
\subsubsection{Threshold Implementations}
Threshold Implementations (TI) were introduced by Nikova et al. in 2006~\cite{nikova2006threshold}, with a goal of reducing the number of refresh gadgets. Previous theoretical work required a refresh gadget for each AND gate of the circuit, and it was vulnerable to glitches. On the other hand, the original TI had higher complexity, but the randomness was only required at the beginning of the circuit while taking glitches into account. TI design is based on secret sharing, threshold cryptography, and multi-party computation protocols. The TI implementation scheme is secure against first-order attacks, as well as higher-order attacks based on comparison of the average power consumption~\cite{nikova2006threshold,reparaz2015note}. This implementation has been extended by Biligen et al. in \cite{bilgin2013efficient,bilgin2012threshold} and expanded in \cite{bilgin2014more,bilgin2015trade,moradi2011pushing,battistello2016horizontal}. This scheme requires inputs to be shared into $dt + 1$ shares, where $d$ is the number of probes used in an attack (so-called attack order), and $t$ is the order of the algebraic function representing the circuit~\cite{gross2017efficient}. 

The main properties of the input shares in TI schemes are correctness, uniformity and non-completeness~\cite{daemen2017changing}. Non-completeness requires that any combination of up to $d$ components of the function (circuit) must be independent of at least one input share~\cite{reparaz2015consolidating}, or in other words that any $d$ component of the function must be independent of all unshared values~\cite{de2018hardware}. Non-completeness requirement provides the security against glitches. An example of TI scheme was given by Reparaz et al. in~\cite{reparaz2015consolidating}, for the function $d=c\oplus{ab}$, for $d=1$ with the following equations
\begin{align}
	d_1&=c_2\oplus{a_2b_2}\oplus{a_1b_2}\oplus{a_2b_1} \\
	d_2&=c_3\oplus{a_3b_3}\oplus{a_3b_2}\oplus{a_2b_3} \\
	d_3&=c_1\oplus{a_1b_1}\oplus{a_1b_3}\oplus{a_3b_1}.	
\end{align}
The number of output shares is greater than the number of input shares for $d>1$. To prevent further increase of the output shares and to avoid glitches, output shares are XORed among themselves after being stored in the registers~\cite{reparaz2015consolidating}. In the example equations above, two functions are cascaded, which means that the output of $c$ is fed as the input into $ab$, by performing re-masking at each step. Re-masking is performed to satisfy the requirement of uniformity, which can also be satisfied by increasing the number of input shares~\cite{de2018hardware, bilgin2015threshold}.

\subsubsection{Consolidated Masking Schemes}
\begin{figure}
  \includegraphics[width=0.5\linewidth]{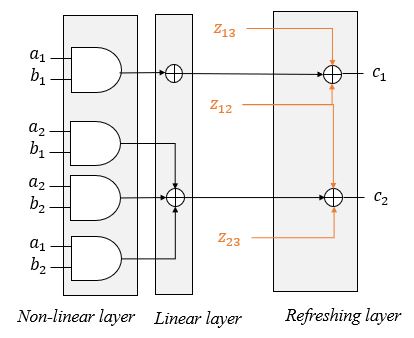}
  \caption{Consolidated Masking Scheme (CMS) AND Gadget}
  \label{fig:CMS}
\end{figure}

Consolidated masking schemes (CMS) were introduced by Reparaz in~\cite{reparaz2015consolidating}. Such schemes use ISW AND and Trichina AND gates as building blocks consolidated with TI. The number of shares required to protect against attacker with $d$ probes is $d + 1$, and compared to TI scheme, the number of input shares is lower. After relating the ISW, TI and Trichina schemes, CMS or generalized scheme was introduced for the higher orders of protection.  
The consolidated masking scheme features all TI properties and also applies the re-masking of ISW to break the correlation between the clock cycles. 
The gates with more than two inputs have been further proposed. CMS has four layers, as shown in Figure ~\ref{fig:CMS}: non-linear, linear, refreshing, and compression layers. The compression layer is used for decreasing the number of output shares. 
The non-linear layer does most of the computation, which performs $a_ib_j$ AND operation on shares, mapping six inputs (3 shares per one input) to nine intermediate values, for $d = 1$. In the linear layer, XOR-ing is performed to compute three output values from nine inputs (which are outputs of the non-linear layer). In a TI scheme, this step must preserve non-completeness. The refreshing layer is then applied on top of the linear layer to consider the re-masking security and characteristic of uniformity. This layer is very important for higher-order security. The final layer is the compression layer which uses XOR gates to decrease the total number of outputs shares when it is greater than the number of shares per individual input.

This scheme was implemented for AES in~\cite{de2016masking}, and compared to TI. It requires more randomness provided by refresh gadgets, but in total occupies less area, specifically, when higher-order security is needed. 

\subsubsection{Domain-Oriented Masking} This is considered as a variant of TI and CMS implementation schemes introduced by Gross et al. in~\cite{gross2016domain,gross2017efficient} and evaluated in~\cite{arribas2018rhythmic}. DOM performs refreshing and compression in two clock cycles~\cite{de2018hardware}, introducing the register (D-flip flop) synchronization prior to the compression stage. Randomness and the number of shares are reduced to $d+1$, but the latency is increased by one clock cycle. This masking implementation scheme is based on two multiplication schemes (AND gadgets) DOM-\textit{indep.} and DOM-\textit{dep.}, which both depend on the SNI requirement. DOM-\textit{indep.} was designed to refresh dependent shared inputs~\cite{moos2019glitch}. Compared to TI masking scheme, DOM is domain-oriented rather than function-oriented, and the minimum number of input shares is used. In this masking implementation, parts of the circuit are considered as a unit (domain) used to develop DOM AND gadgets. In that sense, the domains are individual parts considered independent of each other. For example, if the part of the circuit takes inputs $a_1$ and $b_1$ from domain $1$, all intermediate values (inner-domain terms) calculated with this portion of the circuit are independent of \textit{unshared} inputs $c=a_1+a_2$ and $d=b_1 + b_2$, because $a_2$ and $b_2$ are part of domain $2$.

\begin{figure}
  \includegraphics[width=0.65\linewidth]{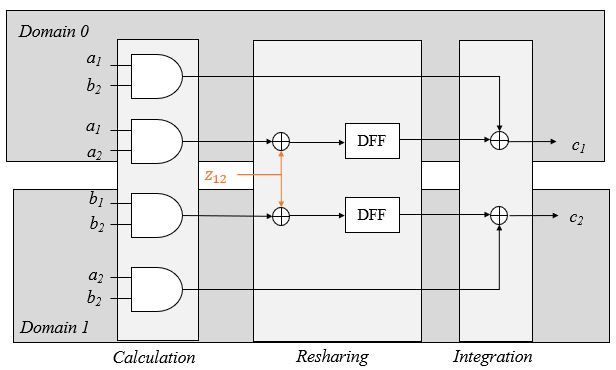}
  \caption{Domain-oriented Masking AND Gadget, where DFF is D-Flip Flop}
  \label{fig:DOM}
\end{figure}

\subsubsection{Unified Masking Approach} 
\begin{figure}
  \includegraphics[width=0.8\linewidth]{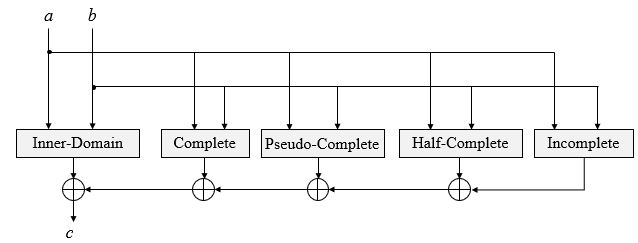}
  \caption{Unified Masking Approach. The AND gate is assembled using complete, pseudo-complete, half-complete, and incomplete blocks.}
  \label{fig:UMA}
\end{figure}

This scheme was introduced by Gross and Mangard in~\cite{gross2018unified}, where for the same level of security as TI schemes, a decrease in the area overhead has been reported. 
Their proposed approach combines software and hardware-based masking methods into a unified one. 
More specifically, it combines the software-oriented parallel masking scheme of Barthe et al.~\cite{barthe2017parallel} with optimized randomness according to~\cite{belaid2016randomness}. 
Although the unified masking approach relies on these study, to extend the previous results achieved for the protection orders $d\leq 4$, Gross and Mangard have suggested to define 4 cases based on the integral multiple of 4. 
These cases are: complete, pseudo-complete, half-complete and incomplete, as shown in Figure~\ref{fig:UMA} and corresponding to $d= 0, 3,2,1\; \text{mod} 4$, respectively. 
To implement the design, various different algorithms and techniques are combined into one, depending on what is needed for the gadget: 

\begin{figure}
  \includegraphics[width=0.5\linewidth]{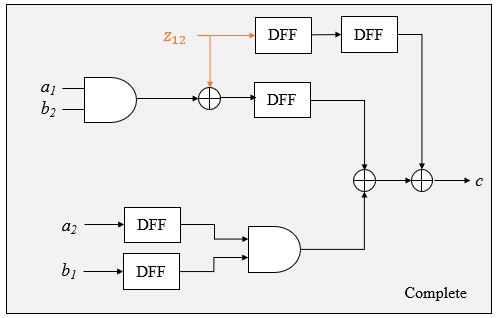}
  \caption{Unified Masking Approach: Complete}
  \label{fig:UMAcomplete}
\end{figure}

\begin{figure}
  \includegraphics[width=0.65\linewidth]{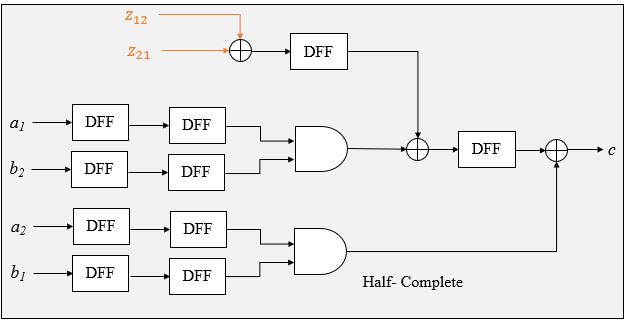}
  \caption{Unified Masking Approach: Half Complete}
  \label{fig:UMAhalfcomplete}
\end{figure}

\begin{figure}
  \includegraphics[width=0.5\linewidth]{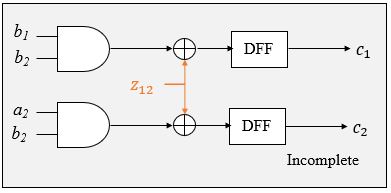}
  \caption{Unified Masking Approach: Incomplete}
  \label{fig:UMAincomplete}
\end{figure} 

\begin{itemize}
  \item Complete and pseudo-complete: Masked AND gate is implemented in parallel according to the scheme in~\cite{barthe2017parallel}. A register is added after each AND gate resulting in, at most, five cycles of latency. 
  \item Half-Complete: For $d=2\;\text{mod} 4$, the randomness optimization is performed~\cite{belaid2016randomness}. For $t\neq2$, DOM is used, see Figure~\ref{fig:UMAhalfcomplete} and note the difference between how the refreshing random bits are used. 
  \item Incomplete: The construction follows the DOM principles. 
\end{itemize}

Each of these cases corresponds to a block used to assemble the AND gate in the unified masking framework. 

\subsubsection{Generic Low-Latency Masking} 
\begin{figure}
  \includegraphics[width=0.38\linewidth]{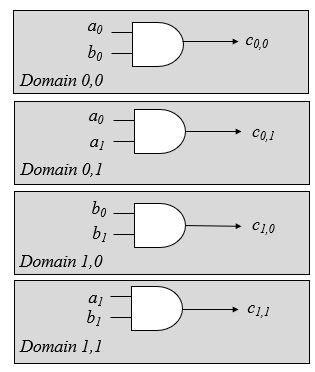}
  \caption{First order low-latency AND Gadget}
  \label{fig:GLM}
\end{figure}

(GLM) is another scheme introduced by Gross et al. in~\cite{gross2018generic},~\ref{fig:GLM}. 
The paper is devoted to the study of the trade-off between randomness usage, chip area and less latency in hardware. 
The proposed generic latency masking is built on the DOM scheme and is applicable to any protection order. 
The main goal of this scheme is to skip the share compression by not requiring fresh randomness during the computation. 
The main observations are that share compression happens purely due to practical reasons, and property of uniformity, and independence across domains is achieved by an increase in number of domains. 

\subsection{Circuit Masking Implementation Summary }

In the literature, there has been a wide range of circuit masking implementations in addition to ones summarized in Table~\ref{tab:tableIMPL}, such as~\cite{bilgin2020low,canright2008very,coron2014higher,goudarzi2018secure}. When it comes to comparing the implementations, different authors report different parameters about their implementations, and Table~\ref{tab:tableIMPL} highlights the common ones, such as Gate-Equivalent Area, security order, number of input shares, and randomness. 
Various implementation schemes already explained have been examined: Threshold Implementation, Consolidated Masking Scheme, Domain Oriented Masking, Unified Masking Approach, Generic Low-Latency Masking, as well as Low-Latency Hardware Masking ~\cite{sasdrich2020low}. The implementations were mostly performed on AES; however some other algorithms were used because of their simpler structure, such as ASCON, KATAN and KECCAK s-boxes. ASCON is used to compare DOM and UMA for non-linear functionality, while KATAN is used on one of the earlier TIs. Randomness is shown to decrease from $td+1$ to $t(t+1)/2$, while the number of input shares remained the same across the work summarized in this table. The area reported varies greatly, as the most recent work reported the area of 3.48kGE~\cite{sasdrich2020low}, which has the smallest footprint. 
\subsection{Security Verification}

The development of attack models has resulted in the possibility of verifying the security of masking schemes and implementations with respect to those abstract attack models and concrete security~\cite{barthe2020improved}.
Bounded moment model, $t$-probing model, and security notions are used to assess the leakage independence of shares, whereas noisy leakage models are used to check the condition of sufficient noise~\cite{barthe2020improved}. However, as already discussed in Section~\ref{sec:attack_models}, the security in the noisy leakage model can be verified by proving security in the probing model. The initial pen-and-paper proofs and implementations were not sufficient as they could easily hide errors,~\cite{coron2007side, coron2013higher, schramm2006higher, rivain2010provably,herbst2006aes}. Furthermore, proving the security of a higher-order masked design could require unrealistically high computing power; hence, splitting the underlying algorithm in smaller gadgets is a useful strategy~\cite{cassiers2020trivially}. 
Adopting this strategy enhances the efficiency of the implementations and allows the composition of secure gadgets. 

\begin{center}
\small
\begin{longtable}[t]{p{1.2cm}|p{1.5cm}|p{1.2cm}|p{1cm}|p{0.7cm}|p{1cm}|p{2cm}|p{0.8cm}}
\caption{Summary of masking schemes implemented on cryptographic algorithms}
\label{tab:tableIMPL}\\\hline
\textbf{Paper} & \textbf{Case Study} & \textbf{Scheme} &\textbf{kGE} & \textbf{$t$} & \textbf{Shares} & \textbf{Randomness} & \textbf{Year}\\
\hline
\cite{bilgin2014higher} & KATAN & TI & N/A & 1,2,3 & $t+1$ & 5 & 2014\\
\hline
\multirow{2}{*}{~\cite{de2016masking}} & \multirow{2}{*}{AES} & \multirow{2}{*}{TI/CMS } & 6.6 & 1 & \multirow{2}{*}{$t+1$} & $54,   (t+1)^2$ & \multirow{2}{*}{2016} \\\cline{4-5} 
&&& 10.4 & 2 && 162\\\hline
\cite{moradi2011pushing} & AES & TI & 11.1 & 1 & $td+1$ & 44 & 2011 \\  \hline
\cite{bilgin2014more, bilgin2015trade} & AES & TI & 8.1 & 1 & $td+1$ & 44 & 2014-5 \\ \hline
\cite{de2015higher} & AES & TI & 7.8 & 2 & $td+1$ & 126 & 2015 \\\hline 
\cite{gross2017efficient} & AES & TI & 6.0 & 1-15 & $t+1$ & $18 t(t+1)/2$ & 2017 \\\hline
\multirow{2}{*}{\cite{gross2016domain}} & \multirow{2}{*}{AES} & \multirow{2}{*}{DOM} & 7.1 & 1-15 & \multirow{2}{*}{$t+1$} & $18 t(t+1)/2$ & \multirow{2}{*}{2016}\\\cline{4-5} \cline{7-7}
&&&5.3&2&&54\\\hline
\multirow{6}{*}{\cite{gross2017reconciling}} & \multirow{6}{*}{ASCON} & \multirow{3}{*}{UMA} & 10.8 & 1 & \multirow{6}{*}{$t+1$} & $5, t(t+1)/4$ & \multirow{6}{*}{2017}\\\cline{4-5} \cline{7-7}
&&&16.4&2&&15\\\cline{4-5}\cline{7-7}
&&&N/R&3-15&&20-320\\\cline{3-5}\cline{7-7}
&&\multirow{2}{*}{DOM} & 10.8 & 1 && $5,t(t+1)/4$ \\\cline{4-5}\cline{7-7} 
&&&16.2&2&&15\\\cline{3-5}\cline{7-7} 
&& UMA &27.2 & 1 & & $320$ \\\hline
\multirow{2}{*}{\cite{gross2018generic}} & \multirow{2}{*}{AES} & GLM 1 & 60.7 & 1 & \multirow{2}{*}{$t+1$} & $2048, 16(t+1)^7$ & \multirow{2}{*}{2018}\\\cline{3-5} \cline{7-7}
&&GLM 2&6.7&1&&$416, 6(t+1)^6 +8(t+1)^2$ \\\hline
\cite{gross2017higher}&KECCAK & DOM&15.7&1&$t+1$&$t(t+1)/2$&2017\\\hline
\cite{sasdrich2020low} & AES& LLHM & 3.48 & 1 & $t+1$ & $36$ & 2020 \\\hline
\end{longtable}
\end{center}

Security verification focuses on the latter (in terms of SNI) ~\cite{barthe2020improved} and global composability. The soundness of the masking scheme has also been shown with respect to well-understood tools, such as leakage detection tool~\cite{reparaz2016detecting}. The goal of security verification is to provide security proof if the circuit is secured. If the circuit is not secure, security verification gives insight into how the circuit can be changed (in terms of randomness, composition, gadgets). 
Security verification research can be divided into two directions: development of security verification frameworks and security verification tools, some of which are summarized in Table~\ref{tab:tableVER}
. 
\subsubsection{Security Verification Frameworks} 
The main goal of verification frameworks is to provide insight into the evaluation and comparison of different implementations coming under attacks by various types of adversaries. 
One of the first studies in this regard was presented by Micali and Reyzin, where analysis of side-channels has been related to the modularity of physically observable computations~\cite{micali2004physically}. 
More specifically, their model relates the notions of modular computation, information leakage, and adversarial power. 
This has been followed by a series of works that focus on defining theoretical models describing SCA and connecting those to more practical approaches to validate the theoretical findings.  
Here we mention a set of such frameworks. 

\begin{center}
\small
\begin{longtable}[t]{p{1cm}|p{2cm}|p{4cm}|p{4cm}}
\caption{Summary of masking verification tools and corresponding techniques}
\label{tab:tableVER}\\\hline
\textbf{Paper} & \textbf{TOOL} & \textbf{Verification Framework} &\textbf{Technique} \\
\hline
\cite{barthe2019maskverif} & MaskVerif & SNI/NI & Matrix Model \\\hline
\cite{coron2018formal} & CheckMasks & Bounded Moment, SNI/NI & Elementary circuit transformation \\\hline
\cite{arribas2018vermi} & VerMI & SNI/NI, Glitch extended probing model & Checking correctness, uniformity, and glitches\\\hline
\cite{arribas2019cryptographic} & VerMI & SNI/NI, Faults, Glitch extended probing model & Extension of VerMI \\\hline
\cite{belaid2018tight} & TightPROVE & Probing Model & Linear algebra (Matrix transformation) \\\hline
\cite{belaid2020tornado} & TightPROVE+, Tornado & Register probing model & Matrix transformation \\\hline
\cite{cassiers2020hardware} & Full verification tool & Robust Probing Model, PINI & Checking for fresh independent randomness, redundant computation of shares, and order of shares not being shuffled. \\\hline
\cite{knichel2020silver} & SILVER & Statistical independence and leakage verification & Reformulation of NI/SNI, t-probing model and avoidance of false negatives \\\hline
\cite{gao2019verifying} & QMSI\textsubscript{INFER} & Quantitative Masking Strength & NI/SNI or SMT solver for interference verifier + QMS calculator\\\hline
\cite{bloem2018formal} & Rebecca Tool & Glitch extended probing model & Fourier coefficients and equation, propagation of input labels\\\hline

\end{longtable}
\end{center}

\vspace{0.5ex}
\noindent \textbf{Unified framework} for side-channel analysis~\cite{standaert2009unified} focuses on the relationship between SCA and information theory. 
In doing so, information-theoretic metrics are defined to compare different implementations from the ``evaluator's'' perspective, who is interested in assessing the security of a scheme independently from the adversary types. 
On the other hand, security metrics have been proposed to quantify a side-channel-based key recovery adversary's success with predefined level of access and computational power. 
In the first category, i.e., information-theoretic metrics for evaluators, the conditional entropy is suggested to argue if an adversary can build a good estimation of the leakage distribution. 
It has also been stated that the information-theoretic metrics and security ones should be combined to evaluate implementations. 
For the security metrics, the adversary's success rate is considered and related to how much the complexity of an exhaustive key search is reduced by launching the SCA. 


\vspace{0.5ex}

\noindent \textbf{Non-interference in probing model}~\cite{belaid2018tight,belaid2017private,barthe2016strong} has been explained in Section~\ref{sec:attack_models}. 
The objective of this framework is to prove the security of the entire implementation by dividing it into smaller gadgets (e.g., code sequences).  
If gadgets fulfill the SNI property, they can be freely composed with other gadgets without violating the SCA resistance. 
This framework mainly discusses the composability and scales up better than its counterpart; however, not all secure masking schemes are directly composable and SNI-compatible, e.g.,~\cite{belaid2016randomness,belaid2017private,gross2017reconciling,barthe2017parallel}. 

\vspace{0.5ex}

\noindent \textbf{Simulatability} is put forward to help proving the security of gadgets~\cite{barthe2015compositional,belaid2016randomness}. 
Simulatability implies that if a set of probes is simulatable, by feeding some input shares to a simulator, simulated probes are delivered that have the same statistical distribution as the true set of probes cf.~\cite{cassiers2020trivially}.  
In other words, the probe-specific simulator outputs the simulated probes without knowing the input shares not given to it (see Section~\ref{sec:attack_models}). 


\vspace{0.5ex}

\noindent \textbf{Statistical Independence and Leakage Verification (SILVER)}~\cite{knichel2020silver} is a verification framework based on statistical independence of probability distributions. 
The security verification relies on the fact that given a targeted security order $d$, at most $d$ independently placed probes can be placed by the attackers. 
The information collected in this way can be combined to reveal the secret. 
Hence, to check the statistical independence, SILVER exhaustively checks all possible adversarial probes and their combination up to $d$ probes. 
This methodology avoids false negative decisions. It reformulates the security notions of NI/SNI/PINI and $t$-probing model in terms of statistical independence of binary random variables.

\vspace{0.5ex}

\noindent \textbf{Notion of masking strength or quantitative masking strength (QMS)}~\cite{eldib2015quantitative,wang2017security,eldib2014qms,wang2019mitigating} quantifies the bias of intermediate values with respect to secret input values and random values. This empirical security notion tells an adversary how many power measurement traces are needed to extract the secret data. When the QMS value is equal to 1, the circuit is perfectly masked~\cite{gao2019verifying}.

\vspace{0.5ex}

\noindent \textbf{VRAPS}, an automatic framework algorithm from~\cite{belaid2020random}, is a security verification tool that considers the random probing model for the sake of capturing horizontal attacks which exploit full leakage traces. However, this tool only works on \textit{small} circuits, i.e., proving the security of gadgets. It quantifies the random probing model security based on the leakage probability. This work also considers the \textit{probing expandability} property, introduced by~\cite{ananth2018private}, as well as SNI security notion.

\vspace{0.5ex}

\noindent \textbf{Multi-model evaluation methodology} introduced in~\cite{journault2017very} is used for bit-sliced algorithms. The methodology relies on probing model \textit{reductions}, in which the confidence in security assessment is built by proving the security in $t$-threshold probing model, followed by bounded moment model, and finally with the noisy leakage model.

\subsubsection{Security Verification Tools} 
As masking-based approaches are devised for \emph{software} and \emph{hardware} implementations, in the same vein, various tools and frameworks have been developed to verify the security of such implementations. 
This differentiation is made based on the observation that for hardware implementations, physical phenomena (e.g., glitches) can affect the side-channel leakage and model, whereas for software ones transitions (e.g., memory transitions) are prominent factors~\cite{barthe2019maskverif}.  
Below, we summarize some of the efforts made to develop verification tools. 

\vspace{10pt}
\noindent\emph{\textbf{Verification of software implementations}}

 \vspace{0.5ex}
 
\noindent \textbf{SMT solvers} worth mentioning are Sleuth~\cite{bayrak2013sleuth}, interference-based SMT solver from~\cite{el2017symbolic}, and SMT solver from~\cite{eldib2014formal,eldib2014smt}, which was the first SMT solver based method to verify the security of the masking countermeasures against side-channel attacks.

\vspace{0.5ex}

\noindent\textbf{TightPROVE} was developed in~\cite{belaid2018tight} based on the framework of SNI/NI security notions from~\cite{barthe2016strong}. The input into the tool is a circuit description, and the output is either a probing security proof or the probing attack. It takes four \textit{security games} to perform the security verification, as introduced in Section~\ref{sec:attack_models} in which attacker plays a computation match against the \textit{simulator}. Probing attack means that the simulator finds the set of wire indices (probed data), which cannot be computed without the knowledge of the plain input. The proposed iterative method is based on linear algebra properties. 

 \vspace{0.5ex}

\noindent \textbf{CheckMasks}~\cite{coron2018formal} is based on elementary circuit transformations. There are two approaches considered in this tool. First, it is an NI/SNI approach, as in~\cite{barthe2017parallel} but in a different programming language, and the second approach is related to the transformation of each gate towards verifying it in NI/SNI security notions. This tool does not consider glitches. 

\vspace{0.5ex}

\noindent \textbf{QMSI\tiny{NFER}} by \cite{gao2019verifying} is used to fill the gap between interference-based tools, which often produce false negatives and SMT solver-based tools that are often very slow and complex. The inputs into this tool are the masked circuit and its inputs, which are either random, public or secret. Inside the tool, there are a masking verifier and QMS calculator. The masking verifier is either interference-based or SMT-based. QMS calculator is applied for those parts of the circuit that are not perfectly masked.

\vspace{0.5ex}

\noindent \textbf{TORNADO} is a tool which extends tightPROVE to tightPROVE\textsuperscript{+}~\cite{belaid2020tornado}, and combines it with bitslicing implementations of cryptographic primitives Usuba~\cite{mercadier2019usuba}. This tool is different from tightPROVE because it considers register-probing model instead of \textit{bit} probing model (\textit{t} probing model). In this case, the leaked information is a vector of multiple values instead of only one. 

\begin{figure}
  \includegraphics[width=\linewidth]{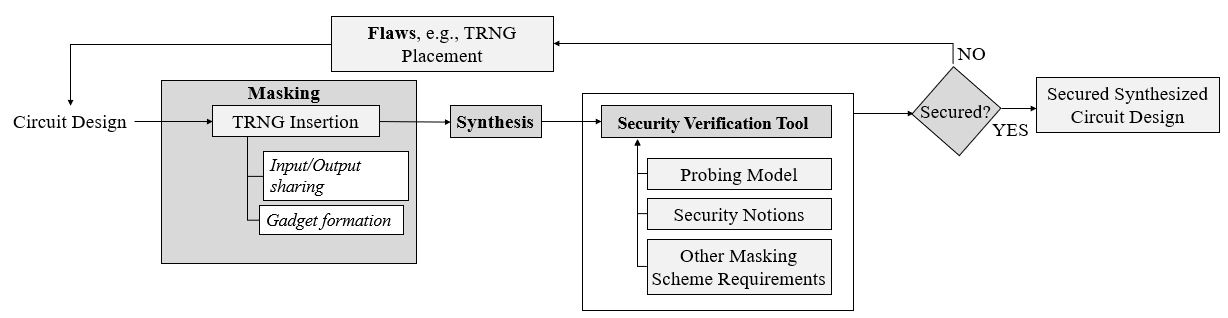}
  \caption{Security verification flow}
  \label{fig:verif}
\end{figure}

\vspace{10pt}
\noindent\emph{\textbf{Verification of hardware implementations}}
\vspace{10pt}

Figure~\ref{fig:verif} illustrates a simplified flow of security verification frameworks. 
Below some of these tools have been discussed. 

 \vspace{0.5ex}

\noindent\textbf{Rebecca}~\cite{bloem2018formal} considers the concept of Fourier coefficient estimation to determine if the circuit is secured. They provide a set of ``weak nets'', if the circuit is not considered secured in $t$-glitch-extended probing model. This tool does not consider advanced notions -- Non-Interference (NI), Strong Non-Interference (SNI), and Probe-Isolating Non-Interference (PINI). It consists of two steps. In the first step, the synthesized Verilog netlist is parsed into a circuit tree using YOSYS tool. Then it is parsed through an SMT Z3 solver, checking for the Fourier coefficients, which are not equal to zero, and possibly leak information. This tool is limited because it does not assess how much information is leaked. 

 \vspace{0.5ex}
 
\noindent \textbf{VerMI} is a verification tool introduced in~\cite{arribas2018vermi} which focuses on the evaluation of TI properties, correctness, uniformity and non-completeness, considering glitches on a synthesized VHDL or Verilog netlist. Non-completeness is checked by first collecting input dependencies of every primary output. Then, the tool checks if all shares of the same variable are included, which results in failing the non-completeness if they are included. It cannot see the difference between the variable and the share, making this task to be very difficult. In addition to non-completeness, uniformity is checked by computing the frequency of output occurrence for multiple groups of input sharing. If the frequency is the same, then the uniformity is preserved. 

\vspace{0.5ex}
 
\noindent\textbf{VerMFi} is a tool built upon VerFI, fault detection tool~\cite{arribas2019cryptographic}, and VerMI~\cite{arribas2018vermi}. In this way, this tool is used for verification of masked implementations while considering fault injection.

 \vspace{0.5ex}
 
\noindent\textbf{Full verification tool}~\cite{cassiers2020hardware} is used as an extension to compiler maskComp, introduced in Section~\ref{sec:masking_schemes}. Compared to maskComp, it considers implementation circuits, compared to abstract circuits consisting of composable gadgets. This tool checks the following assumptions: (1) each gadget is used with fresh independent randomness, (2) there is no redundant computation of shares, and (3) shares' order is not shuffled. This tool, compared to the tool in~\cite{bloem2018formal}, uses YOSYS to develop a circuit tree from the Verilog netlist. The same netlist is also simulated using IcarusVerilog testbench. The circuit tree netlist is then parsed into a graph of physical gadgets, containing \textit{MUX gadgets}, as well as gadgets with inputs that do not hold meaningful value. These two are erased to create a data-flow graph. Verification is performed by checking the security notion property PINI.

\vspace{10pt}
\noindent\emph{\textbf{Verification of hardware and software implementations}}

 \vspace{0.5ex}

\noindent \textbf{MaskVerif}~\cite{barthe2019maskverif} is based on probabilistic non-interference security tightness of observed set of size $t$. This procedure proves probabilistic non-interference~\cite{barthe2015verified}, as well as avoids the \textit{combinatorial explosion} when extracting the values for higher orders. This tool consists of two functions. The first function is used to prove the non-interference property of the observed set. The second function is used to search through all possible observed sets. Security is verified if the second function does not detect any leakage. This function checks if the observed set is independent of secret data. This tool is used for \textit{reasonably} higher orders, considering SNI, NI, and $t$-probing model, and physical defaults glitches. This tool was extended in CheckMask~\cite{coron2018formal} which utilizes more in-depth details. Its concept was also further researched in~\cite{gao2020hybrid}. 

 \vspace{0.5ex}
 
\noindent \textbf{Verification in Fine-Grained Leakage Model}~\cite{barthe2020masking} is one of latest work which focuses on leakages not considered by prior works with respect to physical validation. It is an extension to MaskVerif which performs specific attacks or statistical tests, with emphasis on a Domain-Specific Language called IL.

 \vspace{0.5ex}
 
\noindent\textbf{SILVER tool} based on the SILVER framework considers glitches from robust probing model. It also checks the output uniformity, and reports the weak nets in the circuit. According to~\cite{knichel2020silver}, this tool, which utilizes Reduced Ordered Binary Decision Diagrams (ROBDDs), is the most efficient and effective tool so far.

 


\section{Standardization}
\label{sec:standardization}
One of the first attempts to assess the impact of attacks against cryptosystems or security-critical devices has been conducted under Trusted Computer System Evaluation Criteria (TCSEC)~\cite{TCSEC}. 
It was then followed by Information Technology Security Evaluation Criteria~\cite{ITSEC} and Canadian Trusted Computer Product Evaluation Criteria~\cite{canadian1993canadian}. 
In 1990, these three standards were unified under the umbrella of Common Criteria for Information Technology Security Evaluation (Common Criteria for short). 
According to the Common Criteria, performing an SCA can be categorized as AVA\_VAN, reflecting the resilience of the system under attack, the so-called target of evaluation (TOE). 
In this regard, this resilience is associated with a grade attributed in the AVA\_VAN for a pre-defined threat model, e.g., the level of access to the TOE cf.~\cite{thillard2016countermeasures}. 

Recently, this line has been pursued by the National Institute of Standards and Technology (NIST) to standardize the countermeasures against SCA, namely TI schemes. 
The main standardization process is devoted to the development of \textit{Threshold Schemes for Cryptographic Primitives}, led by NIST,~\cite{brandao2018threshold,brandao2019threshold} for single-device and multi-party settings. The criteria for standardization of TI schemes have been considered in terms of
\begin{enumerate}
  \item \textit{Characterizing features:} threshold, communication interface between environment and components, executing device (single vs. multiple devices), setup and maintenance, 
  \item \textit{Applicability of the scheme:} efficiency of the scheme, applicability to NIST approved crypto primitives, 
  \item  \textit{Implementations:} various platforms, automation,
  \item  \textit{Security:} relation to known SCA, reliability, graceful degradation, security proof with respect to real attacks,
  \item  \textit{Validation:} testability, automated validation, and
  \item  \textit{Licensing, and new standards development:} intellectual property conditions
\end{enumerate}

Characterizing features include types of the threshold, communication interfaces, setup, and maintenance, as well as the executing platform. However, that is not enough for the scheme to be secure in every scenario. Various attack types and models have to be considered, as briefly explained in Sections~\ref{sec:intro} and~\ref{sec:attack_models}. There are four development phases planned: criteria development, call for contribution, evaluation of the contribution, and publishing of the new standards.

The challenge is whether the standards for testing and security verification tools should be created, or if the standards should be created for the design of threshold schemes that \textit{can be tested and verified}. The current roadmap reported in~\cite{brandao2019threshold} has planned the development of threshold schemes for single-device and multi-party domains, where for each domain, there are three routes: thresholdization, compositional design and development of new primitives, as well as the additional route of gadget design. Across the routes, there are various examples, which might have to be considered, such as RSA, ECC key generation, AES circuit design, post-quantum primitives, distributed RNG, etc. This map would also consider four different \textit{modes}: secret-shared key vs. independent multi-signature keys, secret-shared plaintext, fault detection robustness, as well as the asynchronous environment. Standardization of TI does not have to cover all possible scenarios. 

Another NIST standardization that is closely related to circuit masking is the lightweight cryptography standardization process. This process is looking for candidates, which would operate on IoT devices, healthcare, sensor networks, and similar mobile devices. Most of the current crypto schemes have been developed for desktop and servers, and they are not the most optimized for the networked web of devices. There has been work reported on evaluating the masking schemes on NIST lightweight candidates with a less complex s-box function than AES and RSA. According to~\cite{delooking}, many of the NIST candidates have not considered physical attacks for their development, or they claimed that due to the simple s-box, and abundance of research on AES, the masking is applicable to their work. However, that is not sufficient to claim resistance against physical attacks. 

\section{Lessons Learned and an Outlook for Future Studies}
\label{sec:lessons}

Side-channel attacks and their corresponding countermeasures have received particular attention over the years. 
When surveying studies in this well-established, we have observed a gap between theoretical understanding and practical applications and implementations. 
Below we summarize some of our observations. 

\vspace{0.5ex}

\noindent\textbf{Requirement for Knowledge of the Field: }As a prime example, one can mention the subtle difference between the \emph{privacy} and \emph{$t$-probing security} models (see Section~\ref{sec:attacks_probing_model}). 
To understand such a difference, a hardware security expert must have a good knowledge of theoretical results and their interpretation; otherwise, the implementation would not fulfill the requirements. Consequently, it would be less secure/insecure than what the theory has indicated. 
This is definitely a tedious task since even the terminology used in theoretical works could be inconsistent, i.e., the same phenomena or notions is known under different names, e.g., $t$-NI security is equivalent to the $t$-probing security as defined in~\cite{carlet2015algebraic}, and perfect $t$-probing security as defined in~\cite{coron2013higher,rivain2010provably} (see Section~\ref{sec:side_channel_attacks}).  

\vspace{0.5ex}

\noindent\textbf{Assumptions Made in Theoretical Studies: }
One of the reasons why masking has been widely adopted is the easy understanding of its security requirements. 
In fact, masking schemes can be thought of as instances performing computations on secret-shared data~\cite{faust2018composable}. 
For this, the two main assumptions are: (1) each leakage (sample) depends on a limited number of shares (ideally one) or their linear combination of~\cite{barthe2017parallel}, and (2) the leakages of the shares are noisy enough. 
If these hold, masking ensures that the measurement complexity of any side-channel attack increases exponentially in the number of shares, whereas the implementation cost of that could exhibit a polynomial increase in the number of shares (almost quadratically~\cite{faust2018composable}). 

Nevertheless, in practice, the assumptions on independence and noisiness are not always met. 
For instance,~\cite{coron2013higher} has shown an example of attack leveraging the combination of $d' < d$ shares to extract sensitive information (against the assumption on the independence) cf.~\cite{faust2018composable}. 
Another example is a decrease in the security order due to the physical defaults including glitches (i.e., combinatorial recombinations)~\cite{mangard2005side,mangard2005successfully}, transition-based leakages (i.e., memory recombinations)~\cite{coron2012conversion,balasch2014cost}, and  couplings (i.e., routing recombinations)~\cite{de2017does} cf.~\cite{faust2018composable}. 
In addition to the above two issues, the condition defined on the basis of the physical noise is not easy to fulfill. 
This condition considers two factors: (1) the physical noise, generally assumed almost equal for all the operations, and (2) the number of operations. 
Horizontal attacks, however, have demonstrated that an adversary can efficiently exploit the multiple leakages of the operations made available due to an increase in the number of shares~\cite{battistello2016horizontal,grosso2018masking}. 
Interestingly, in~\cite{grosso2018masking}, it is concluded that a higher level of physical noise is necessary to ensure security. 
To estimate or determine this level, an information-theoretic analysis is provided that relates the required noise level to the number of exploitable operations~\cite{grosso2018masking}. 

\vspace{0.5ex}

\noindent\textbf{Software Implementations vs. Hardware Ones: }
Another interesting point is the difference between software and hardware implementations. 
The first point to be made in this regard is the invalidity of the assumptions underlying the concept of both hardware-oriented and software-oriented masking. 
For software schemes, due to the Hamming distance leakage usually visible in memory-element transitions, the assumptions may not hold~\cite{balasch2014cost}. 
On the other hand, for hardware implementations, glitches have been exploited as a source of leakage~\cite{mangard2005side,mangard2006pinpointing}. 
Steps within the consolidations framework have already been taken to show the connection between circuit constructions in these two domains~\cite{barthe2017parallel}. 
Moreover, the fundamental difference regarding the serial and parallel manipulations of the shares has been addressed by introducing leakage models capturing the inherent characteristics of hardware implementations~\cite{barthe2017parallel}. 

Nonetheless, even for such a hardware-oriented scheme, several aspects of implementation, for instance, randomness requirement and optimization, remain as important and somehow unresolved issues as discussed in the literature~\cite{gross2018unified}. 
Such issues typically have various facets, for example, if the amount of required randomness is reduced, more registers are needed, and consequently, more latency would be introduced to the design~\cite{gross2018generic}. 
Therefore, to design a hardware masking scheme, the possible trade-offs should be considered carefully. 

\vspace{0.5ex}

\noindent\textbf{Extension of Existing Models: } 
One of the most pressing needs is for probing models and tools to assist hardware masking designers in designing the primitives. 
As an example, the noisy leakage models were believed to be realistic, although the assumptions made about the noise and its parameter could be invalid cf.~\cite{prest2019unifying}. 
Within these frameworks, the design parameters, e.g., noise parameter, should be under the control of the designer~\cite{prouff2013masking}, which could not be possible in practice. 

Besides this shortcoming, probing or even noisy leakage models should provide a basis for more realistic attacks extracting information from circuits. 
One of the requirements is to take into account the capability of attackers that have access to devices facilitating the SCA. 
Yao et al.~\cite{yao2018fault} and Papadimitriou et al.~\cite{papadimitriou2020you} have presented fault-assisted attacks, where it could be challenging to formulate the attacker's ability in leakage and probing models. 
Another attack that is beyond the scope of the probing model has been introduced in~\cite{krachenfels2020real}. 
These studies further highlight the need for extended models. 

\vspace{0.5ex}

\noindent\textbf{Summary: }
Various physical implementation issues have been thought of in the literature, and accordingly, steps in the design of masking schemes have been proposed. 
However, we believe that a broad range of hardware security experts can bring together the experience and knowledge needed to augment the theoretical development of circuit masking schemes. 
This would eventually create robust, effective, and implementable masking countermeasures. Therefore, hardware security community researchers and practitioners are needed to bring this circuit masking theory into practice. 

\subsection{Future Directions: Use of Circuit Masking in Hardware Security Community}

Another goal of this survey is to discuss the future directions for hardware security community, and how the knowledge presented in this paper can be used, as we discuss in the following subsections. 
\textit{Note that for many of these applications below, a critical first step would be to extend masking tools and concepts beyond cryptographic circuits so that they are suitable for more general digital circuits.}

\subsubsection{Gate Level Information Flow Tracking}
Information flow tracking (IFT) is a technique used in computer science to pinpoint how information moves through a computing system. 
Hardware IFT is a variant that specifically targets the unintentional (e.g., design flaws, timing side channels~\cite{oberg2014leveraging}, etc.) and malicious (e.g., hardware Trojans~\cite{hu2016detecting}) information leakage that leads to vulnerabilities in IC designs~\cite{hu2021hardware}.
Principles from the masking community such as non-interference ($t$-NI/SNI,PINI), along with their associated verification tools, might be applicable (with some adaptation) to hardware IFT. 
For example, by using increasing the number of probes in such tools, it may be possible to analyze if multiple, simultaneous information flows in a circuit lead to information leakage. 
Further, the analysis of glitches and couplings between signals, which has been performed in the circuit masking community and seemingly overlooked in hardware IFT, can be explored. 



\subsubsection{IP Protection for AI-enabled Hardware}
Traditionally, masking techniques have been proposed to protect software and hardware implementations of cryptographic systems. 
These systems, however, are not the only targets for side-channel attacks nowadays.  
The new targets are AI-enabled hardware. 
SCA can lead to several concerns regarding the security of AI hardware due to its applications in numerous fields, including autonomous cars,
smartphones, and robots -- just to name a few. 
Hardware platforms specialized for AI embody AI IP cores and assets stored on these platforms (for a survey on the assets, see~\cite{tajik2020artificial}). 
Hence, side-channel attacks and their countermeasures have become immensely important and discussed extensively in the literature~\cite{batina2019csi,wei2018know,hua2018reverse,dubey2019maskednet}. 

\subsubsection{Hardware Trojan Detection} 
A hardware Trojan is a malicious modification of an IC, which may be performed by untrusted IP vendors, CAD tools or SoC designers, and foundries~\cite{xiao2016hardware}. 
As discussed above, circuit masking properties and verification tools might be useful for detecting information flows caused by hardware Trojans. 
In addition, while hardware Trojan trigger selection/insertion has been studied quite often in the literature (e.g.,~\cite{wei2013undetectable,shakya2017benchmarking,cruz2018automated}), intelligent hardware Trojan payload selection has been overlooked. For example, rather than a Trojan directly leaking a cryptography key-bit to an output (something that hardware IFT and/or formal methods would surely detect), multi-probe analysis could be used to find a set of internal nets whose combination leaks the key.


\subsubsection{Evaluation of Probing Sensors and Physical Countermeasures}
There is a disconnect between the theoretical probing leakage models of the masking community and the practical probing~\cite{wang2017probing} countermeasures described in the literature. 
For example, probing sensors~\cite{weiner2019calibratable} are only applied to buses, while active shields/meshes are inefficiently used to protect the entire IC. Since probing models and tools can identify nets of interest for an attacker to probe, they can be used to distribute these physical countermeasures in a more efficient way.
~\cite{wang2019physical} investigates a CAD flow to automatically place-and-route shields over the most critical nets, but its approach for selecting the nets is ad hoc and inapplicable to masked circuits.

\subsubsection{Quantification of Side Channel-based Disassembly}

In order to identify malware, instruction-level side-channel monitors, also called side-channel
disassemblers (SCDs), have been proposed. SCDs can be used in coarse and fine modes. For the former, power consumption
and/or EM radiation of a running device can be measured and statistically compared to the same signals from a device running malware-free code. For the latter, the instruction at each line of code is reverse engineered from the side channel measurements for forensic purposes~\cite{park2019leveraging}. The effectiveness of SCDs can be improved by incorporating probing models into the disassemblers. If the goal is to prevent side channel-based reverse engineering, the models may be used to determine the series of instructions with the least amount of leakage.


\subsubsection{Hardware Weakness/Vulnerability Standardization Efforts} 

Besides the efforts of NIST, which are specific to SCA and threshold-based masking schemes, multiple groups are working on cataloging the broader set of hardware weaknesses, vulnerabilities, scoring metrics, and countermeasures in order to help individuals and organizations understand how these
weaknesses occur in real-world systems and how to mitigate them. The most notable is the software security ecosystem that includes the Common Weakness Enumeration (CWE) and Common Vulnerabilities and Exposures (CVE) developing by MITRE, which is now being extended to hardware by the Hardware CWE Special Interest Group (SIG)~\cite{CWE_SIG}. Another example is the Accellera Systems Initiative's IP Security Assurance Working Group~\cite{Accellera_IP}. These efforts are works in progress, which often acknowledge side-channel leakage and countermeasures such as masking but remain vague. They can significantly benefit from the efforts by the cryptographic community to categorize masking schemes, develop gadgets, and quantify leakage. 

\subsection{Future Directions: Enrich Circuit Masking with Hardware Security Approaches}

Another goal of this survey is to discuss which approaches outside of circuit masking techniques could be combined or integrated with masking to provide better protection against physical probing and side-channel attacks.

\subsubsection{Incorporation of Physical Models} 

The provable information leakage models developed by the cryptographic community abstract out complex physical phenomena and/or rely on inaccurate assumptions (e.g., independence of shares). This is why many provably secure schemes are still vulnerable to side-channel attacks once implemented in silicon. Recent work in the hardware security community has focused on developing tools and methodologies to capture and understand the impact of the physical sources of leakage \textit{at design time} such that they can be mitigated before chip fabrication. For example,~\cite{lin2020fast} proposes scalable approaches for evaluating the location-dependent side-channel leakage within the on-chip power distribution network (PDN). In addition,~\cite{vsijavcic2020sweeping} finds that parasitics in the PDN break independence assumptions. Such results can be useful for either refining the current models and assumptions or creating physical design tools/methodologies (e.g., decoupling)  that maintain assumptions.



\subsubsection{Probing Sensors and Physical Countermeasures}
Physical countermeasures, such as active shields, can protect against probing attacks. Those shields can specifically be placed to protect the points of possible information leakage, known to designers with leakage detection tool. In this way, the weakest points of circuit masking would be additionally protected. In the same way, probing sensors could be placed to protect the weak points. However, the disadvantage here is that if the sensor is only connected to the weak point, then once removed with bypass attack, the attacker has direct access to the leakage point of the circuit. 
Designing an effective sensor-integrated masked scheme is, therefore, an interesting direction to pursue.  


\section{Conclusion}
\label{sec:conclusion}

This survey has demonstrated the great need for leakage detection methodology, masking countermeasures, and verification.  
When surveying the literature, we have made key observations: 
(1) the gap between theory and practice can be bridged if the adversary models are defined more precisely to reflect the real-world attack scenarios, (2) such models should either be generic to take into account properties of both hardware and software implementations or make a clear distinction between the requirement for each of those implementations, (3) to make the models and tools more practical and to bring them closer to the hardware security designers and practitioners, methods and measures must be developed and become more accessible to hardware engineers. 
Our survey has aimed to address the letter and pave the way for hardware security experts. 

\begin{acks}
This work was supported in part by NSF under Project 1717392, in part by SRC under Grant 2769.001, and in part by AFOSR MURI under Project FA9550-14-1-0351.
\end{acks}

\bibliographystyle{ACM-Reference-Format}
\small
\bibliography{references}

%
%
%

\end{document}